\documentclass[11pt,letterpaper]{scrartcl}
\usepackage{setspace}
\usepackage{mathpazo} 
\usepackage[scaled]{helvet} 
\usepackage{courier} 
\normalfont
\usepackage[T1]{fontenc}

\usepackage{amsmath}
\usepackage{amssymb}

\usepackage[T1]{fontenc}
\usepackage[utf8]{inputenc}
\usepackage{slashed}
\usepackage[parfill]{parskip}
\usepackage{siunitx}
\usepackage{csquotes}
\usepackage[hidelinks]{hyperref}
\usepackage{cleveref}
\usepackage{todonotes}
\usepackage{scalefnt}
\usepackage{microtype}

\usepackage[affil-it]{authblk}

\DeclareOldFontCommand{\rm}{\normalfont\rmfamily}{\mathrm}
\DeclareOldFontCommand{\sf}{\normalfont\sffamily}{\mathsf}
\DeclareOldFontCommand{\tt}{\normalfont\ttfamily}{\mathtt}
\DeclareOldFontCommand{\bf}{\normalfont\bfseries}{\mathbf}
\DeclareOldFontCommand{\it}{\normalfont\itshape}{\mathit}
\DeclareOldFontCommand{\sl}{\normalfont\slshape}{\@nomath\sl}
\DeclareOldFontCommand{\sc}{\normalfont\scshape}{\@nomath\sc}
\DeclareRobustCommand*\cal{\@fontswitch\relax\mathcal}
\DeclareRobustCommand*\mit{\@fontswitch\relax\mathnormal}

\newcommand{\abbrev}{\scalefont{.9}}

\newcommand{\NNLO}{\text{\abbrev NNLO}}
\newcommand{\NLO}{\text{\abbrev NLO}}
\newcommand{\LO}{\text{\abbrev LO}}
\newcommand{\EFT}{\text{\abbrev EFT}}
\newcommand{\FT}{\text{\abbrev FT}}
\newcommand{\SM}{\text{\abbrev SM}}
\newcommand{\BSM}{\text{\abbrev BSM}}
\newcommand{\IR}{\text{\abbrev IR}}
\newcommand{\QCD}{\text{\abbrev QCD}}
\newcommand{\PDF}{\text{\abbrev PDF}}
\newcommand{\LHC}{\text{\abbrev LHC}}
\newcommand{\NLOs}{\text{\abbrev NLO$^*$}}
\newcommand{\NLOd}{\text{\abbrev NLO$^\dagger$}}
\newcommand{\MCFM}{\text{\abbrev MCFM}}

\def\MC@NLO{{\sc MC@NLO}}

\def\PYTHIA8{{\sc PYTHIA8}}
\def\HERWIG++{{\sc HERWIG++}}

\def\spa#1.#2{\langle #1  #2\rangle}
\def\spab#1.#2.#3{\langle #1|#2|#3]}
\def\spb#1.#2{[ #1 #2]}

\def\beq{\begin{equation}}
\def\eeq{\end{equation}}
\def\beqn{\begin{eqnarray}}
\def\eeqn{\end{eqnarray}}

\newcommand{\abs}[1]{\lvert#1\rvert}

\allowdisplaybreaks[4]

\begin{document}

\title{The Higgs boson at high $p_T$}

\author{Tobias Neumann%
	\thanks{Electronic address: \texttt{tobiasne@buffalo.edu}} }

\author{Ciaran Williams%
	\thanks{Electronic address: \texttt{ciaranwi@buffalo.edu}}}

\affil{Department of Physics, University at Buffalo\\ The State University of New York, Buffalo 14260 USA}

\date{}

\newcommand{\zero}{{(0)}}
\newcommand{\one}{{(1)}}
\newcommand{\two}{{(2)}}
\newcommand{\ztwo}{\zeta_2}
\newcommand{\zthree}{\zeta_3}
\newcommand{\cf}{C_F}
\newcommand{\ca}{C_A}
\newcommand{\nf}{n_f}
\newcommand{\cfs}{C_F^2}
\newcommand{\Tcm}{\tau_{cm}}

\maketitle

\begin{abstract} We present a calculation of $H+j$ at \NLO{} including the effect of a finite top-mass. Where possible we include the complete 
	dependence on $m_t$. This includes the leading order amplitude, the infrared poles of the two-loop amplitude and the real radiation amplitude. The remaining 
	finite piece of the virtual correction is considered in an asymptotic expansion in $m_t$, which is accurate to $m_t^{-4}$. By successively including more $m_t$-exact pieces, the dependence on the asymptotic series diminishes and we find convergent behavior for $p_T^H>m_t$ for the first time. Our results justify rescaling by the $m_t$-exact \LO{} cross section to model top-mass effects in \EFT{} results up to $p_T$ of 250 to \SI{300}{GeV}. We show that the error made by using the \LO{} rescaling becomes comparable to the \NNLO{} scale uncertainty for such large energies. We implement our results into the Monte Carlo code {\abbrev MCFM}.
\end{abstract}

\clearpage
\tableofcontents

\section{Introduction}

The standout result from the first run of the \LHC{} was the discovery of a Higgs boson~\cite{Chatrchyan:2012xdj,Aad:2012tfa}. The continued exploration of the physics 
associated with the Higgs defines one of the key goals of the current run of the machine (Run II). During Run II a much larger data 
set will be accumulated than that used to initially discover, and study, the Higgs. Accordingly current and future analyses will be able to significantly 
extend our understanding of the Higgs, and hence the mechanism by which the electroweak symmetry is broken.

Studies of the Higgs boson during Run I were primarily limited to its inclusive properties~\cite{Aad:2015lha,ATLAS:2015bea,Aad:2015mxa,Khachatryan:2014jba,Aad:2015gba,Giardino:2013bma}. Over a range of production and decay channels good agreement ($\sim\text{$10$--$20$\%})$ with the predictions of the Standard Model (\SM{}) was found. With the successful 
completion of the {\abbrev N$^{3}$LO} inclusive Higgs cross section~\cite{Anastasiou:2015ema,Anastasiou:2016cez} the theoretical errors associated with these measurements should be significantly reduced. 
In addition to inclusive studies, a major advantage of Run II is the ability to study the Higgs differentially with far greater accuracy than that obtained in Run I \cite{Aad:2015lha,Aad:2016lvc,Aad:2015tna,Aad:2014tca,Aad:2014lwa,Khachatryan:2015yvw,Khachatryan:2015rxa,CMS:2015obs}.

Apart from the intrinsic interest in understanding the only known fundamental scalar, there are strong motivations to understand the Higgs as a tool to discover 
or constrain Beyond the Standard Model (\BSM{}) physics. The Higgs boson may not be exactly as predicted by the \SM{}: \BSM{} physics could occur, for instance, 
through extended Higgs sectors (which are possibly in an alignment limit such that the \SI{125}{GeV} Higgs is \SM{}-like), or through anomalous interactions induced by heavy new 
particles. The Higgs may couple directly to \BSM{} particles such as top-partners or through portal interactions to scalars associated with additional sectors. Many of these situations are best constrained by investigating the Higgs at higher transverse energies~\cite{Harlander:2013oja,Dawson:2014ora,Dawson:2015gka,Banfi:2013yoa,Azatov:2013xha,Grojean:2013nya,Schlaffer:2014osa,Buschmann:2014twa,Buschmann:2014sia,Langenegger:2015lra,Ghosh:2014wxa}, where the \BSM{} effects can become more apparent than in the inclusive result. 

Current precision results for {\abbrev N$^{3}$LO} Higgs inclusive~\cite{Anastasiou:2015ema,Anastasiou:2016cez},
\NNLO{} Higgs plus jet~\cite{Boughezal:2013uia,Chen:2014gva,Boughezal:2015dra,Boughezal:2015aha} and \NLO{} Higgs plus multi-jets~\cite{Campbell:2006xx,Campbell:2010cz,Cullen:2013saa} 
are available in the Effective Field Theory (\EFT{}) limit in which the top quark is integrated out. This is because the technical complexity associated with including the exact top quark loop is significantly greater than for the corresponding calculation in the \EFT{}. For the Higgs inclusive cross section, the exact top mass dependence is known at \NLO{}~\cite{Graudenz:1992pv,Harlander:2005rq,Anastasiou:2006hc}. Although no exact result is available, the top-mass effects for inclusive Higgs production at \NNLO{} have been computed using an asymptotic expansion~\cite{Harlander:2009mq,Pak:2009dg,Harlander:2009bw,Pak:2009bx,Harlander:2016hcx} and matched  to the high energy limit \cite{Marzani:2008ih,Marzani:2008az,Harlander:2009my,DelDuca:2003ba}. As a result, the effect of the top-mass was estimated to be at the percentage level on the inclusive cross section.
For the Higgs plus one-jet topology currently only \LO{} is known exactly~\cite{Baur:1989cm,Ellis:1987xu}. Additionally, \LO{} predictions are available in the full theory for Higgs plus two~\cite{DelDuca:2001eu,DelDuca:2001fn} and three~\cite{Campanario:2013mga,Greiner:2016awe} jets. Combining the various jet topologies with the exact top mass dependence into a matched parton shower framework has also been studied recently~\cite{Frederix:2016cnl,Buschmann:2014sia}. While the analytic calculation of $2\rightarrow 2$ two-loop amplitudes including the top mass and an external massive boson remains a formidable challenge, exciting recent results using numerical methods to evaluate the two-loop master integrals have been employed to calculate di-Higgs production including the exact top mass at \NLO{}~\cite{Borowka:2016ypz,Borowka:2016ehy}. Finally, top mass effects using the high energy behavior for Higgs plus jet have also recently been studied~ \cite{Forte:2015gve,Caola:2016upw}.

We therefore find ourselves in a rather unappealing situation in that, while the \EFT{} works extremely well for inclusive rates and differential predictions at low scales, differential studies which probe scales $\ge  m_t$ are exposed to large theoretical uncertainties due to the low perturbative order of the full theory results. To estimate the impact of higher order corrections in the full theory, an asymptotic series in $m_t^{-2}$ was computed \cite{Neumann:2014nha,Harlander:2012hf}. For Higgs transverse momenta $p_T$ up to scales $\simeq m_t$ finite top mass effects could be estimated to be at the few percent level. However, in the region of larger $p_T$, which we are primarily interested in here, the series rapidly diverged. The principal aim of this paper is to significantly extend the range of validity of the asymptotic series approach, allowing for finite top-mass predictions and more general statements regarding the top mass effects to be inferred. This can be achieved by using, where possible, the full top-mass dependent amplitudes. Given that the process is at \NLO{}, the infrared (\IR{}) poles of the two-loop amplitude (and associated finite pieces) are already known. We will therefore minimize the dependence on the asymptotic series by only using it to compute the finite part of the two-loop virtual amplitude. Our results will be made available in the Monte Carlo code {\abbrev MCFM}
~\cite{Campbell:1999ah,Campbell:2011bn,Campbell:2015qma}.

Our paper proceeds as follows: In section~\ref{sec:calculation} we discuss the setup of our calculation and detail the exact dependence on the asymptotic series. In section~\ref{sec:results} we present our results and compare them to various approximations currently used in the literature. Finally, in section~\ref{sec:conc} we present our conclusions.

\section{Calculation}\label{sec:calculation}

\begin{figure}
\begin{center}
\includegraphics[width=\textwidth]{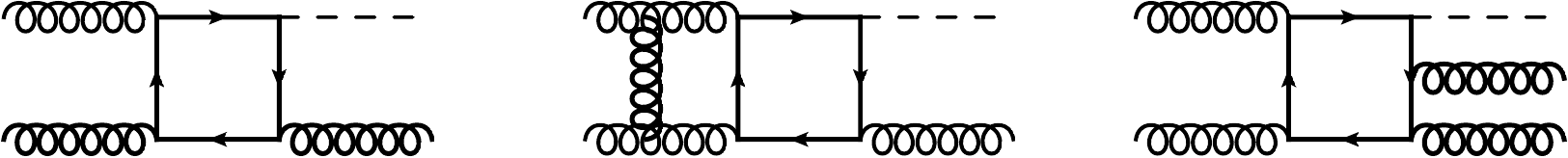}
\caption{Representative Feynman diagrams for the production of a Higgs boson plus one jet at LO (left) and NLO (center, right). 
The NLO corrections include two-loop \enquote{virtual} topologies (center), and one-loop Higgs plus two parton \enquote{real} topologies (right).}
\label{fig:feynd}
\end{center}
\end{figure}

In this section we discuss the technical details of our calculation and introduce different approximations that have been used to model finite top-mass effects.
For illustrative purposes, we show representative Feynman diagrams for the process under consideration in \cref{fig:feynd}. 
Including the effect of the top-mass in the full theory, the \LO{} calculation corresponds to that of a one-loop amplitude. Given the 
relative simplicity of this calculation, results have been known for some time~\cite{Ellis:1987xu} (see also refs.~\cite{Mantler:2009,Keung:2009bs} for a modern discussion, from which we take the \LO{} amplitudes and tensor structures \cite{Mantler:2009,Harlander:2012pb}). The \NLO{} corrections to this process, represented by diagrams such as those in the center 
and right hand side of \cref{fig:feynd}, are considerably more complicated. The real corrections mandate the inclusion of the Higgs plus four parton amplitudes. 
This calculation was first performed in ref.~\cite{DelDuca:2001eu,DelDuca:2001fn}, studying the top mass effects on Higgs plus two jet production. For the \NLO{} calculation, the presence of the second parton as a jet is not required, and the single unresolved limit of this amplitude is readily explored. Therefore, the computation of this amplitude requires care to ensure numerical stability. 

The two-loop virtual corrections, illustrated in the center of the figure, are the most technically challenging part of the calculation. Techniques required to reduce the system to master integrals which can be analytically solved are not yet mature enough to handle tasks of this complexity. 
However, recent progress using numerical methods to evaluate similar master integrals for di-Higgs productions have recently produced phenomenologically usable results~\cite{Borowka:2016ypz,Borowka:2016ehy}. On the analytic side an analytic \NLO{} calculation of $H\to Z\gamma$ also marks some progress \cite{Bonciani:2015eua}.

Since our calculation is performed at \NLO{}, we can write the two-loop amplitude as
\begin{align}
\mathcal{A}^{(2)}_j(m_t,m_H,s,t,u) = 
\mathcal{I}^j_1(\epsilon,s,t,u) \mathcal{A}^{(1)}_j(m_t,m_H,s,t,u) + \mathcal{F}^{in}_j(m_t,m_H,s,t,u)\,.
\end{align}
In the above equation $s$, $t$ and $u$ are the Mandelstam variables associated with the partonic kinematics, $m_t$ and $m_H$ are the masses of the top quark and Higgs boson, respectively. The partonic structure of the amplitude is defined by $j$, where $j=g$ refers to the $gggH$ amplitude and $j=q$ to $q\overline{q}gH$. In terms of these quantities the above equation states that the two-loop virtual amplitude can be written as a function $\mathcal{I}^j_1(\epsilon,s,t,u)$ which multiples the one-loop amplitude and a second function $\mathcal{F}^{in}_j(m_t,m_H,s,t,u)$ defining the remaining pieces. Crucially, all the \IR{} poles are contained in the $\mathcal{I}_1$ piece, such that $\mathcal{F}^{in}_j$ is finite. This \IR{} divergent piece can be obtained from the general structure of \QCD{} amplitudes, \cite{Catani:1996vz}. Therefore, the unknown part of the two-loop virtual amplitude corresponds to the $\mathcal{F}^{in}_j$ term.

Given that $\mathcal{F}^{in}_j$ is currently unknown, it is worth discussing the various approximations one could employ to attempt to quantify the effect of the top quark mass. 
Since the \EFT{} results are known to an impressive \NNLO{} accuracy, a natural thing to do is to use the \EFT{} result rescaled by the ratio of the \LO{} full theory result to the \LO{} \EFT{} result, that is
\begin{align}
\sigma^{\NNLO}_{\EFT-\text{rescaled}} = \sigma^{\NNLO}_{\EFT} \times \frac{\sigma^{\LO}_{\FT}}{\sigma^{\LO}_{\EFT}}\,.
\end{align} 
This rescaling will always work at the level of the total cross section, but complications arise when more differential quantities are considered. In general it is not possible to perform this rescaling fully-differentially, since a higher order phase space in the \EFT{} requires a mapping back to the Born phase space of the rescaling. However, after completing the calculation one can rescale an observable by the ratio 
\begin{align}
\frac{\mathrm{d}\sigma^{\NNLO}_{\EFT-{\text{rescaled}}}}{\mathrm{d} \mathcal{O}} = \frac{\mathrm{d} \sigma^{\NNLO}_{\EFT}}{\mathrm{d}\mathcal{O}} \times \frac{\left(\mathrm{d} \sigma^{\LO}_{\FT}/ {\mathrm{d}\mathcal{O}}\right)}{\left( \mathrm{d}\sigma^{\LO}_{\EFT}/{\mathrm{d} \mathcal{O}}\right)}\label{eqn:loreweight}\,.
\end{align} 
This allows one, for instance, to obtain top mass improved $p_T^H$ or $y_H$ distributions. Such an approach was pursued recently \cite{Chen:2016zka}. 
This methodology has two potential drawbacks: Firstly, the ratio may not be defined over the full range of the observable phase space. For example if a jet is required in the final state, the region $p_T^H < p_T^{\text{jet}}$ is non-existent at \LO{}. Therefore, in this region no rescaling can be defined, and a discontinuity in the observable definition is introduced at the phase space boundary. Secondly, the precision of the \NNLO{} calculation is impressive, and one may naturally worry that the \LO{} nature of the rescaling introduces a comparable uncertainty given its low order in the perturbative expansion. Therefore, a much more appealing rescaling would be 
\begin{align}
\frac{\mathrm{d}\sigma^{\NNLO}_{\EFT-\text{rescaled}}}{\mathrm{d} \mathcal{O}} = \frac{\mathrm{d} \sigma^{\NNLO}_{\EFT}}{\mathrm{d} \mathcal{O}} \times \frac{\left(\mathrm{d} \sigma^{\NLO}_{\FT}/ \mathrm{d}\mathcal{O}\right)}{\left(\mathrm{d}\sigma^{\NLO}_{\EFT}/\mathrm{d}\mathcal{O}\right)}\,.
\end{align} 
The study of this modified rescaling, and its comparison to the equivalent LO definition is a principal aim of this paper. 

Aside from rescaling the \EFT{}, one could attempt to compute the top mass effects as accurately as possible with existing tools. At high $p_T^H$, one expects contributions from $H+2j$ to be significant. This piece can be computed exactly in the top mass and, therefore, one can set up a calculation which includes these contributions: they constitute the real part of the calculation, such that only the virtual part needs to be approximated.
Clearly in order to have a sensible prediction all poles must cancel. This can be done for instance by rescaling the virtual pieces, by the \LO{} ratio. Since the virtual and \LO{} part share the same phase space, this can be performed pointwise. After rescaling, the \IR{} poles cancel and one effectively defines 
\begin{align}
\mathcal{F}^{\text{in}}_j(m_t,m_H,s,t,u) =  \mathcal{F}^{\text{in}}_{j,\EFT}(m_H,s,t,u) \frac{\mathcal{A}^{(1)}_j(m_t,m_H,s,t,u) }{\mathcal{A}^{(1)}_{j,\EFT}(m_H,s,t,u) }\,.
\label{eq:nlodag}
\end{align}
Such an approach was undertaken in ref.~\cite{Frederix:2016cnl}.

An alternative approach to rescaling is to attempt to quantify the top mass effects in $\mathcal{F}^{\text{in}}_j$ approximately with an asymptotic series expansion~\cite{Harlander:1999cs,Smirnov:2002pj} in $\Lambda/m_t$, where $\Lambda$ can be any kinematical scale of the process. This approach has the advantage that it quantifies the impact of the top mass directly at \NLO{}, but suffers from the disadvantage that such series typically diverge at energy scales $\Lambda$ comparable to the top quark mass $m_t$. In this approach one has some freedom in whether the amplitude or the matrix element gets expanded asymptotically. 
Specifically we define two terms, 
\begin{align}
\mathcal{F}^{{\text{in}}}_{j,\text{RI}} =  2 \left[{\text{Re}} \left(\mathcal{F}^{\text{in}}_j(m_t,m_H,s,t,u)\mathcal{A}^{(1)}_j(m_t,m_H,s,t,u)^*\right) \right]_{\text{asy}}\label{eqn:virtfullasy}
\end{align}
and
\begin{align}
\mathcal{F}^{{\text{in}}}_{j,\text{SI}} =  2 {\text{Re}} \left(\left[\mathcal{F}^{\text{in}}_j(m_t,m_H,s,t,u)\right]_{\text{asy}}\mathcal{A}^{(1)}_j(m_t,m_H,s,t,u)^*\right)\label{eqn:virtmixed}\,.
\end{align}
In the above equations, $[\ ]_{\text{asy}}$ refers to the terms which are asymptotically expanded. In the RI term the full interference is asymptotically expanded, whereas in the SI term only the IR finite parts of the two-loop amplitude are asymptotically expanded.

\subsection{Calculation Details }

The necessary ingredients for our calculation can be split into two groups corresponding to the real and virtual parts of the calculation. We calculated the Higgs plus four parton loop amplitudes which fully include the effect of the top quark using unitarity techniques outlined in refs.~\cite{Badger:2008cm,Mastrolia:2009dr}\footnote{We used the spinor helicity library S@M \cite{Maitre:2007jq} frequently.}. To validate our analytical calculation, we compared it to results obtained using an in-house implementation of the $D$-dimensional unitarity algorithm presented in ref.~\cite{Ellis:2008ir}.
 
The asymptotic expansion \cite{Smirnov:1994tg,Harlander:1999cs,Smirnov:2002pj} of the two-loop virtual corrections has been performed with the exp/q2e \cite{Harlander:1997zb,Seidensticker:1999bb,Steinhauser:2000ry} codes: the massive one- and two-loop tadpoles were computed with MATAD \cite{Steinhauser:2000ry} and massless one-loop integrals were reduced to scalar master integrals with Reduze \cite{vonManteuffel:2012np}\footnote{We would like to thank Tom Zirke for gluing code between exp/q2e and Reduze.}. To extract the finite piece from the asymptotically expanded result, we subtract the relevant asymptotic expansion of $\mathcal{I}^j_1(\epsilon,s,t,u)\, \mathcal{A}^{(1)}_j(m_t,m_H,s,t,u)$. The \IR{} poles are then restored in the full theory by replacing these terms with their exact counterparts. 
 
The \IR{} singularities associated with the real and virtual pieces are regularized using Catani-Seymour dipole subtraction~\cite{Catani:1996vz}. Both the integrated dipoles and real subtraction pieces are defined with the full top mass dependence. The massive scalar master integrals, needed for the evaluation of the matrix elements, are computed with {\abbrev QCDLoop} \cite{Ellis:2007qk,Carrazza:2016gav}.  All parts are assembled in the {\abbrev MCFM} framework. For singular regions in the real emission we dynamically switch between double and quad precision.

 \section{Results}\label{sec:results}

The principal aims of our study are to significantly extend the range of validity of the asymptotic series approach, allowing for finite top-mass predictions, and to investigate the validity of using born rescaling schemes as indicated in \cref{sec:calculation}. In the first subsection we compare our improved predictions to the full asymptotic series as previously calculated in the literature. The foundation of our study is based on Higgs $p_T$ distributions, where the Higgs can be produced either inclusively or with associated jet requirements. Concerns regarding missing threshold effects in the asymptotic series are also addressed. Subsequently we introduce different \EFT{} rescaling approaches taken in the literature and evaluate their validity. Finally we present some Higgs+jet phenomenology using our recommended predictions.

Our input parameters are $m_H=\SI{125}{\GeV}$ for the Higgs boson mass, $m_t=\SI{173.5}{\GeV}$ for the on-shell top-quark mass, and $\mu_R=\mu_F=\sqrt{m_H^2+p_{T,H}^2}$ for a common renormalization and factorization scale. We use {\abbrev{CT14}} \PDF{}s \cite{Dulat:2015mca} at \NLO{} accuracy for the \NLO{} cross sections and at \LO{} accuracy for the \LO{} cross sections. Except for the Higgs inclusive cross sections, and unless specified otherwise, we use the anti-$k_T$ jet algorithm with $p_{t,\text{jet}}^\text{min}=\SI{30}{\GeV}$, $\abs{\eta_{jet}^\text{max}}=5$ and $R=0.5$.

In order to be as conservative as possible we use a center of mass energy $\sqrt{s}$ of $\SI{14}{\TeV}$ which we consider to be the worst case scenario with respect to finite top-mass effects for the asymptotic series. This is because the expansion is effectively in $\Lambda/m_t$, and the kinematical scales $\Lambda$ grow with $\sqrt{s}$.
Further, the type of experimental analysis which will be particularly sensitive to the high $p_T^H$ region will require a large amount of data. Such a large data set is most likely to be accumulated during prolonged runs at the maximal design energy of $\SI{14}{\TeV}$.

\subsection{Dependence on the asymptotic series}

\begin{figure}
	\centering\includegraphics[width=\textwidth]{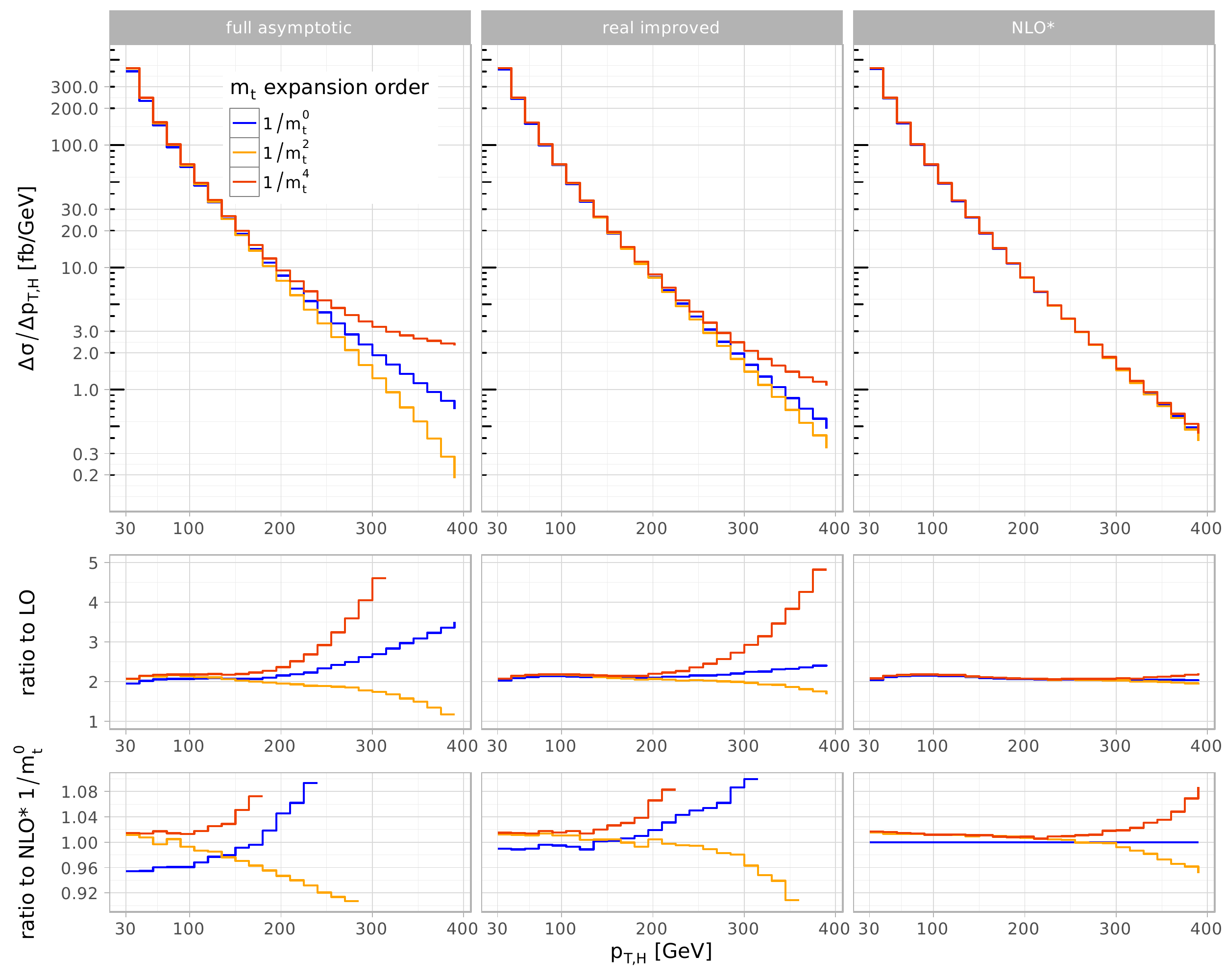}
	\caption{Higgs inclusive $p_T$ spectrum for three different approximations, each taking into account higher orders of an asymptotic expansion in $1/m_t$. The upper panel shows the absolute distribution, while the lower two panels display the ratio to the \LO{} distribution and the \NLOs{} $1/m_t^0$ approximation, respectively. }
	\label{fig:hptcalc}	
\end{figure}

Our first results, for the inclusive Higgs $p_T$ spectrum, are presented in~\cref{fig:hptcalc}. Since the predictions are inclusive, no jet 
requirement is imposed, we simply require a Higgs boson with $p_T^H \ge \SI{30}{GeV}$. Shown in~\cref{fig:hptcalc} are three successive 
predictions which treat the correction in different approximations. From left to right the number of terms which are asymptotically expanded 
diminishes. In the leftmost panel the calculation is fully in the asymptotic series (and corresponds to a re-calculation of the existing literature result~\cite{Neumann:2014nha,Harlander:2012hf}). 
The dependence on the asymptotic series is readily apparent. The $m_t^{-2}$ and $m_t^{-4}$ predictions rapidly diverge from each other, indicating poor convergence in the asymptotic series and an inability to accurately quantify the impact of the top mass. 

The second (central) plot in~\cref{fig:hptcalc} displays the (new) prediction obtained with the real pieces computed exactly in the top mass, and the virtual pieces computed in the RI formalism described in section~\ref{sec:calculation}, which we recall corresponds to the full interference being asymptotically expanded. It is immediately clear that, when compared to the full asymptotic expansion discussed previously, the dependence on the order in the asymptotic expansion is dramatically reduced.  An additional pleasing feature apparent in this plot is made clear upon inspection of the lower panel, which presents the ratio of the $m_t^{-2}$ and $m_t^{-4}$ series to the $m_t^0$ term. This ratio, in essence, determines the impact of the next order in the asymptotic expansion. At low $p_T^H$ the correction going from $m_t^0$ to $m_t^{-2}$ is around a few percent. However, the inclusion of the next term in the asymptotic expansion does not significantly change the result in the region $p_T^H < \SI{100}{GeV}$. Beyond this scale the predictions again begin to differ significantly from one another. 
However, we postulate that in the region $p_T^H < \SI{100}{GeV}$ the closeness of the two curves indicates a converging asymptotic series, and thus deviations from the exact top mass prediction should be essentially zero. A potential fly in the ointment arises from threshold effects, which are not modeled by the asymptotic series. We will address this issue shortly, but for now we use the smallness of the difference between the $m_t^{-2}$ and $m_t^{-4}$ terms to define a region of convergence. Inside this region we postulate that our prediction will accurately capture the top mass effects. For the RI prediction in the central panel, this corresponds to $p_T^H < 100$ GeV.

The final (right) panel in~\cref{fig:hptcalc} presents the $SI$ prediction defined in section~\ref{sec:calculation}. For this prediction, only the amplitude for the finite part of the two-loop virtual is obtained in the asymptotic series. The improvement from the central panel is clear, the differences between the $m_t^{-2}$ and $m_t^{-4}$ are even smaller. The region of convergence is significantly extended from around \SI{125}{GeV} for the RI prediction to around \SI{250}{GeV}. Crucially, this is the first prediction which extends the region of convergence beyond $p_T^H > m_t$ GeV. We therefore have obtained a prediction which captures the top mass corrections in a stable way beyond the scale in which the \EFT{} breaks down. This prediction therefore represents our best prediction in terms of the small dependence of the order of the asymptotic series. Even in the largest bin the dependence is around 8\%, which is still smaller than the \NLO{} scale variation (around 20\% which we study shortly). Going forward, we refer to this prediction as \NLOs{}.

\subsection{Threshold effects}

\begin{figure}
	\centering\includegraphics[width=0.8\textwidth]{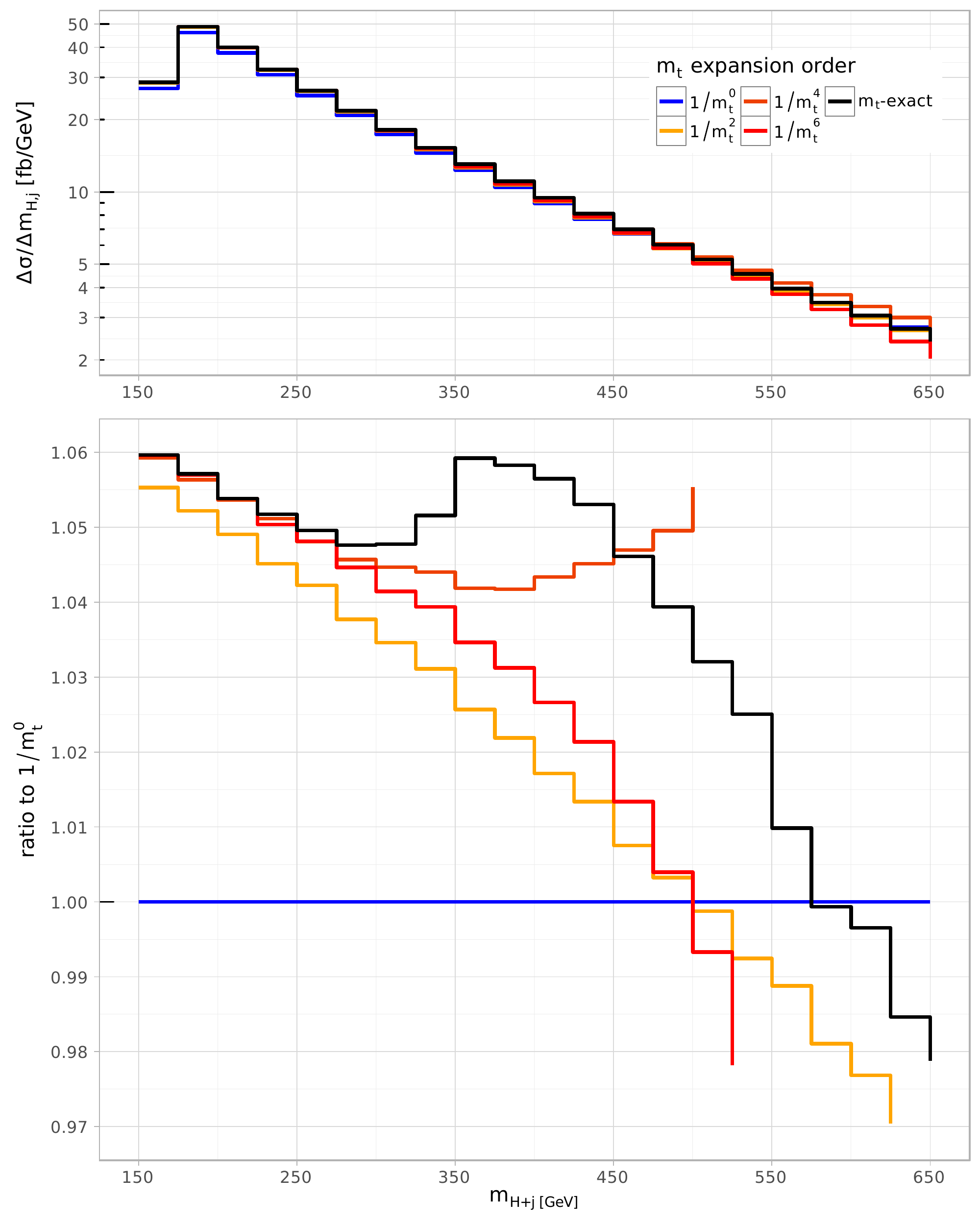}
	\caption{Invariant mass spectrum of the Higgs plus hardest jet system at \LO{}. The upper part displays the absolute distribution, while the lower part displays the ratio to the \EFT{} result.}
	\label{fig:invmassLO}
\end{figure}

As briefly mentioned in the previous section, there could be non-negligible threshold effects that are not described by the asymptotic expansion in the convergent region. To address this issue we study the invariant mass spectrum of the Higgs plus hardest jet system $\Delta\sigma/\Delta m_{H,j}$.\footnote{As a reminder, we use the anti-$k_T$ jet algorithm with $p_{t,\text{jet}}^\text{min}=\SI{30}{\GeV}$, $\abs{\eta_{jet}^\text{max}}=5$ and $R=0.5$.}
This distribution is shown in \cref{fig:invmassLO} at \LO{}. Since the threshold effects are rather small, they are not visible in the absolute distribution in the upper panel of the figure. Instead one has to look at the ratio to the \EFT{}, which contains no finite top-mass effects at all, and thus no threshold effects. In this way any threshold effect will be visible as an increased cross section at $m_{H,j}\simeq 2m_t$. As evident from the plot, the asymptotic expansion correctly converges to the $m_t$-exact result up to the threshold region beginning at $\sim\SI{300}{GeV}$. The one-loop box diagrams lead to a clean threshold effect.

At \NLO{} we take into account the full top-mass dependence of the born piece, the real radiation and all infrared associated pieces of the virtual corrections, as well as the born part of the virtual corrections. The only part where threshold effects are missing are the finite parts of the two-loop integrals entering the virtual corrections, since they are taken in the asymptotic expansion. 

\begin{figure}
	\centering\includegraphics[width=0.9\textwidth]{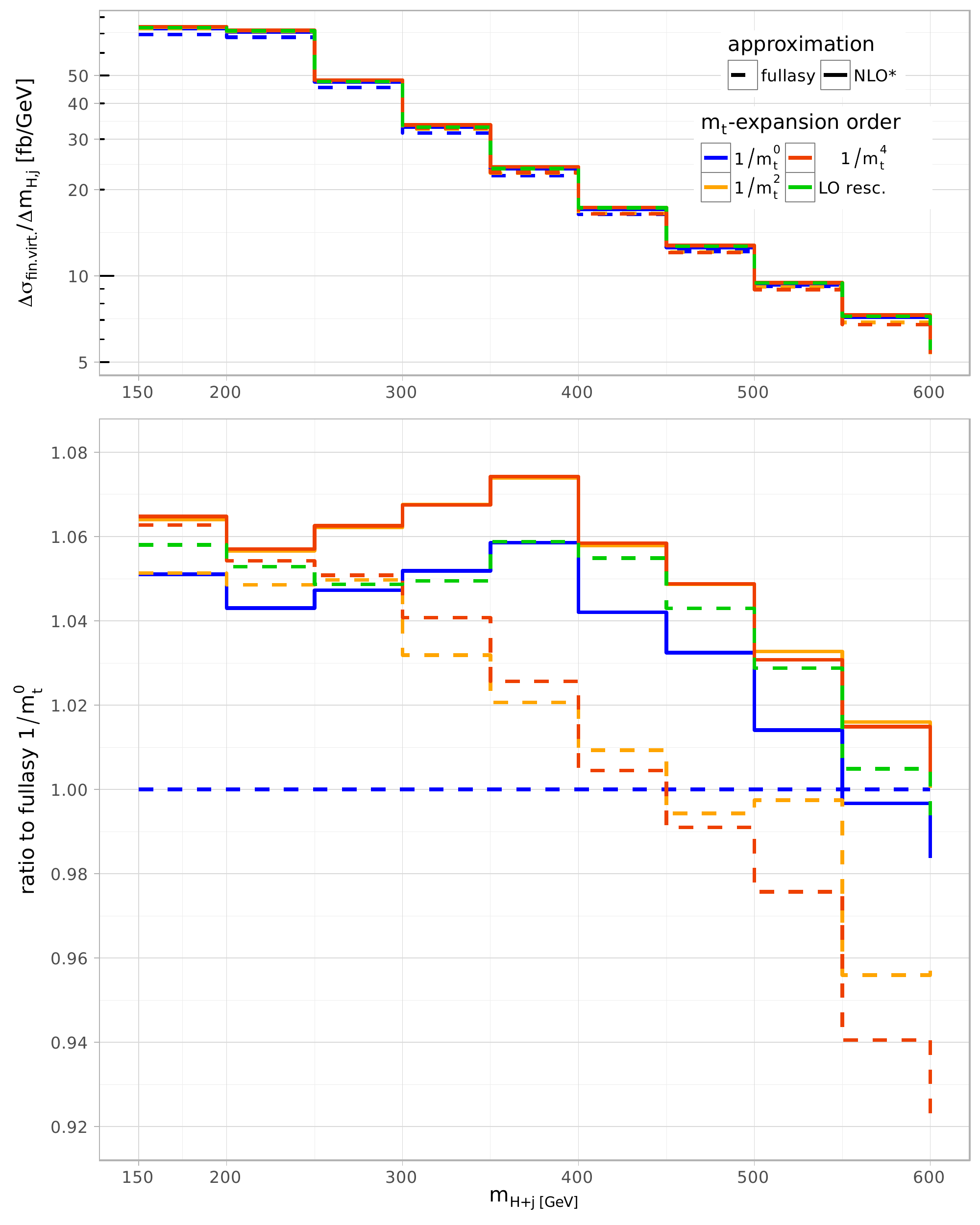}
	\caption{Invariant mass spectrum of the Higgs plus hardest jet system at \NLO{}. The upper part displays the absolute distribution, while the lower part displays the ratio to the \EFT{} result.}
	\label{fig:invmassNLO}
\end{figure}
We show the \NLOs{} $\Delta\sigma/\Delta m_{H,j}$ distribution in \cref{fig:invmassNLO}.
Again, in terms of the absolute distribution, as shown in the upper panel, the threshold effects are too small to be visible. Therefore, we consider the ratio to the \EFT{} result in the lower panel. With solid lines we show our \NLOs{} results, whereas with dashed lines the results of the full asymptotic expansion are displayed. The green dashed curve shows the \LO{}/\LO{}-\EFT{} rescaled \NLO{}-\EFT{} result. Since latter is used as a normalization, the green dashed curve in the lower panel coincides with the $m_t$-exact black curve in the lower panel of \cref{fig:invmassLO}.

Firstly, let us note that we see convergence of the asymptotic series for our \NLOs{} result over the whole mass range considered. As expected, since we take into account all those $m_t$-exact pieces as stated previously, we see an enhancement in the region of $\simeq\SI{350}{GeV}$. It is also evident that the full asymptotic expansion does not model the threshold effects at all. Our improved result however does show features of a threshold. This alone does not allow us to draw conclusions about the threshold effects we miss through the finite piece of the two-loop integrals, but the comparison with the \LO{} rescaled \EFT{} result does: The \LO{} rescaled \EFT{} result shows threshold effects of a similar magnitude and shape. We also note that the real radiation has a significant impact on the observable, softening the spectrum. In summary, we are satisfied that we capture the dominant threshold effects and that any deviations induced by the missing finite two-loop virtual amplitude should be minor.

\subsection{EFT rescaling approaches}

In this subsection we compare various \EFT{} rescaling approaches to our improved \NLOs{} results.
We investigate the difference with respect to an overall rescaling of the virtual \EFT{} piece as outlined in \cref{eq:nlodag} (for notational convenience we refer to this as \NLOd{}) and consider the ratios $\NLOs{} / \text{\NLO{}-\EFT{}}$, $\NLOd{} / \text{\NLO{}-\EFT{}}$ and $\LO{} / \text{\LO{}-\EFT{}}$.

\begin{figure}
	\centering\includegraphics[width=0.75\textwidth]{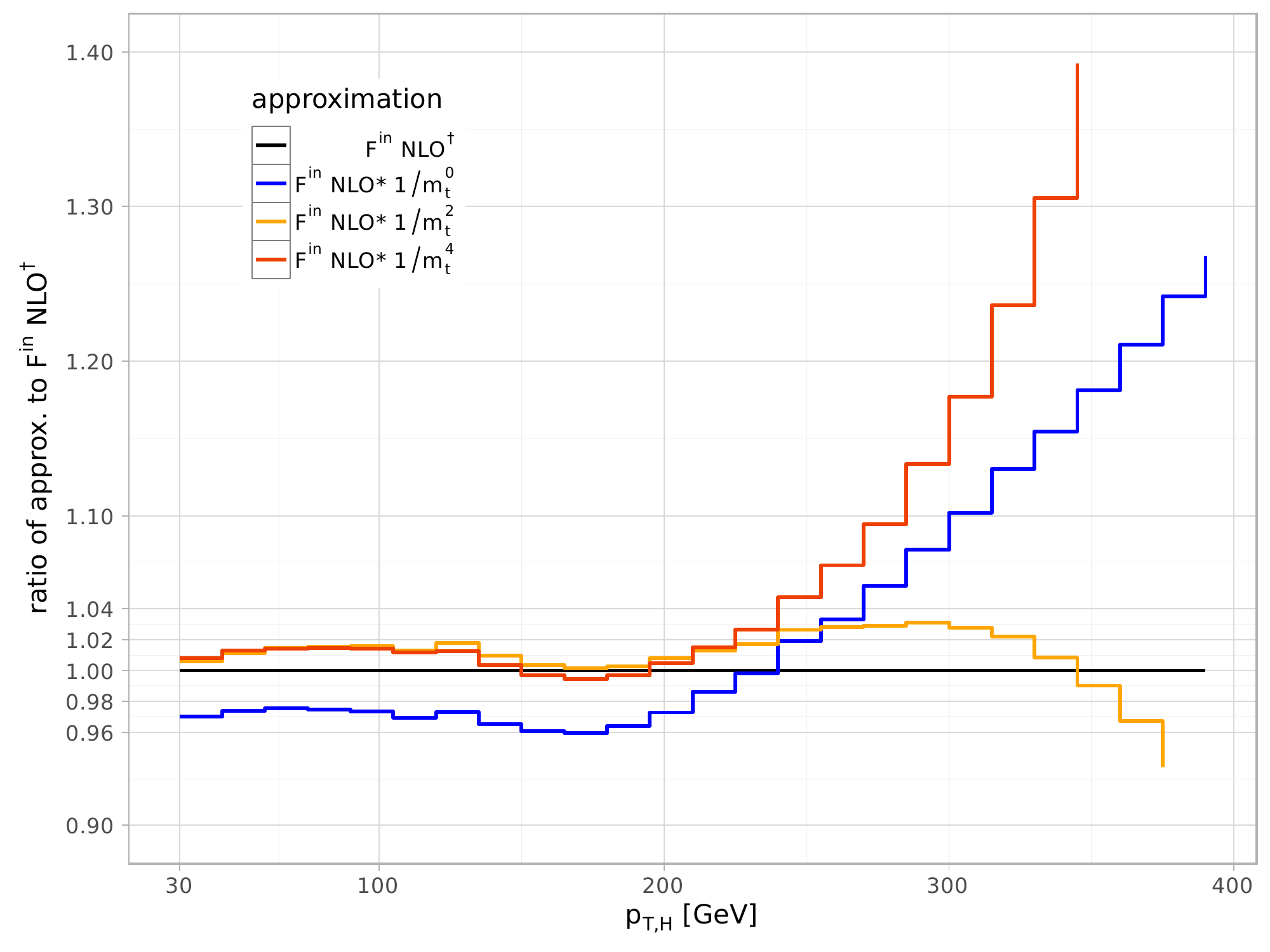}
	\caption{\label{fig:virtcomparison} Ratio of the \NLOs{} finite virtual piece $\mathcal{F}^\text{in}_{\text{SI}}$ in the asymptotic expansion as in \cref{eqn:virtmixed} to the \LO{} rescaled \EFT{} virtual piece $\mathcal{F}^\text{in}\,\NLOd{}$ as in \cref{eq:nlodag}.}
\end{figure}

Since \NLOs{} and \NLOd{} only differ by their treatment of the finite virtual piece (c.f. \cref{eq:nlodag,eqn:virtfullasy,eqn:virtmixed}), we therefore focus on the difference in this piece only for now. Our results are presented in \cref{fig:virtcomparison} as a ratio of our \NLOs{} predictions to the \NLOd{} one. For Higgs $p_T$ less than $\simeq\SI{225}{GeV}$ the difference stays below 2\%. Given that previously the rescaling employed by the \NLOd{} prediction was an uncontrolled approximation, our new results validate this rescaling in the region of asymptotic series convergence.

\begin{figure}
	\centering\includegraphics[width=0.9\textwidth]{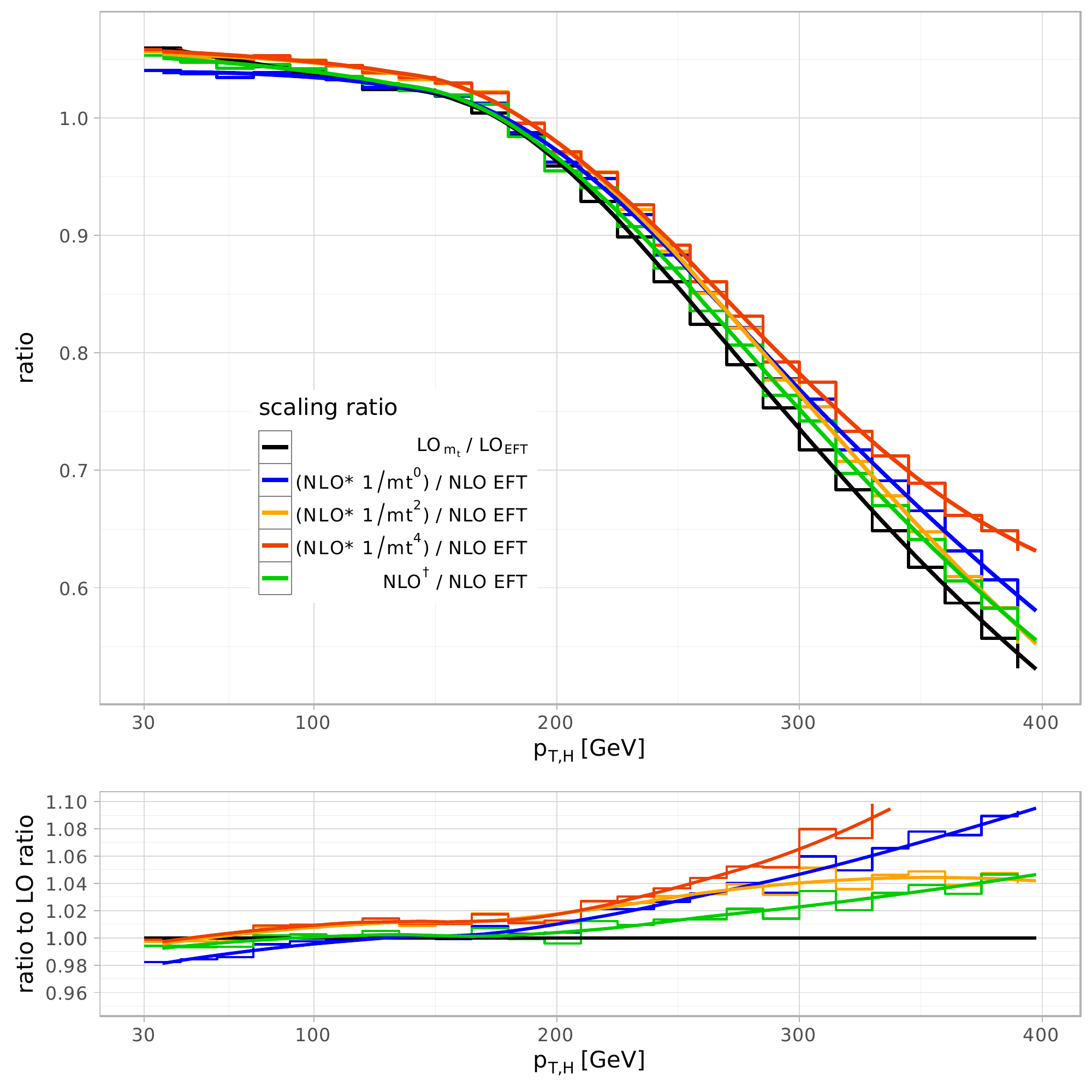}
	\caption{\label{fig:nnloratios}Rescaling ratios for the inclusion of mass effects in the \NNLO{} Higgs inclusive $p_T$ spectrum. The upper panel shows ratios directly, while the lower panel displays the ratio to the \LO{} ratio to give an estimation of the error made by using the \LO{} rescaling scheme. The solid lines are locally weighted scatterplot smoothing curves to guide the eye.}	
\end{figure}

We now turn our attention to the \EFT{} rescaling options. For example in ref.~\cite{Chen:2016zka} the $\LO{}/\text{\LO{}-\EFT{}}$ rescaling has been used to take into account top-mass effects at \NNLO{}. With our improved \NLOs{} result we are able to justify this by studying the \NLO{} rescaling \NLOs{}/\NLO{}-\EFT{}. We examine the various rescaling ratios in \cref{fig:nnloratios} for the Higgs inclusive $p_T$ spectrum. In the upper panel the ratios are shown directly, whereas in the lower panel the ratio of the \NLOs{} and \NLOd{} rescaling ratios to the \LO{} rescaling ratio are given. Any deviation from one indicates an error by just using the \LO{} rescaling method to include top-mass effects. Up to $\simeq\SI{225}{GeV}$ the asymptotic series converges well and the ratios differ by less than three percent. Beyond \SI{250}{GeV} the series clearly begins to diverge.

Note that although at \SI{300}{GeV} the dependence on the asymptotic series reaches a few percent, the overall trend is to provide a somewhat larger rescaling ratio than the \LO{} prediction. At \SI{300}{GeV} an increase of at least $3\text{-}4\%$ is indicated, being comparable to the \NNLO{} scale uncertainty of $\simeq8\%$ \cite{Chen:2016zka,Boughezal:2015aha,Boughezal:2013uia}. This is in agreement with the \NLOd{} approach, which also suggests a reweighting leading to a harder $p_T$ spectrum. 

Our recommendation is to take into account the $\mathcal{O}(m_t^{-2})$ terms for our \NLOs{} prediction. They capture the finite top-mass effects for low to medium $p_T$ where the asymptotic series converges, and mimic the \NLOd{} approximation toward higher $p_T$. This will be the default setting for the upcoming \MCFM{} release that includes Higgs+jet at \NLOs{} accuracy.

\begin{figure}
	\centering\includegraphics[width=\textwidth]{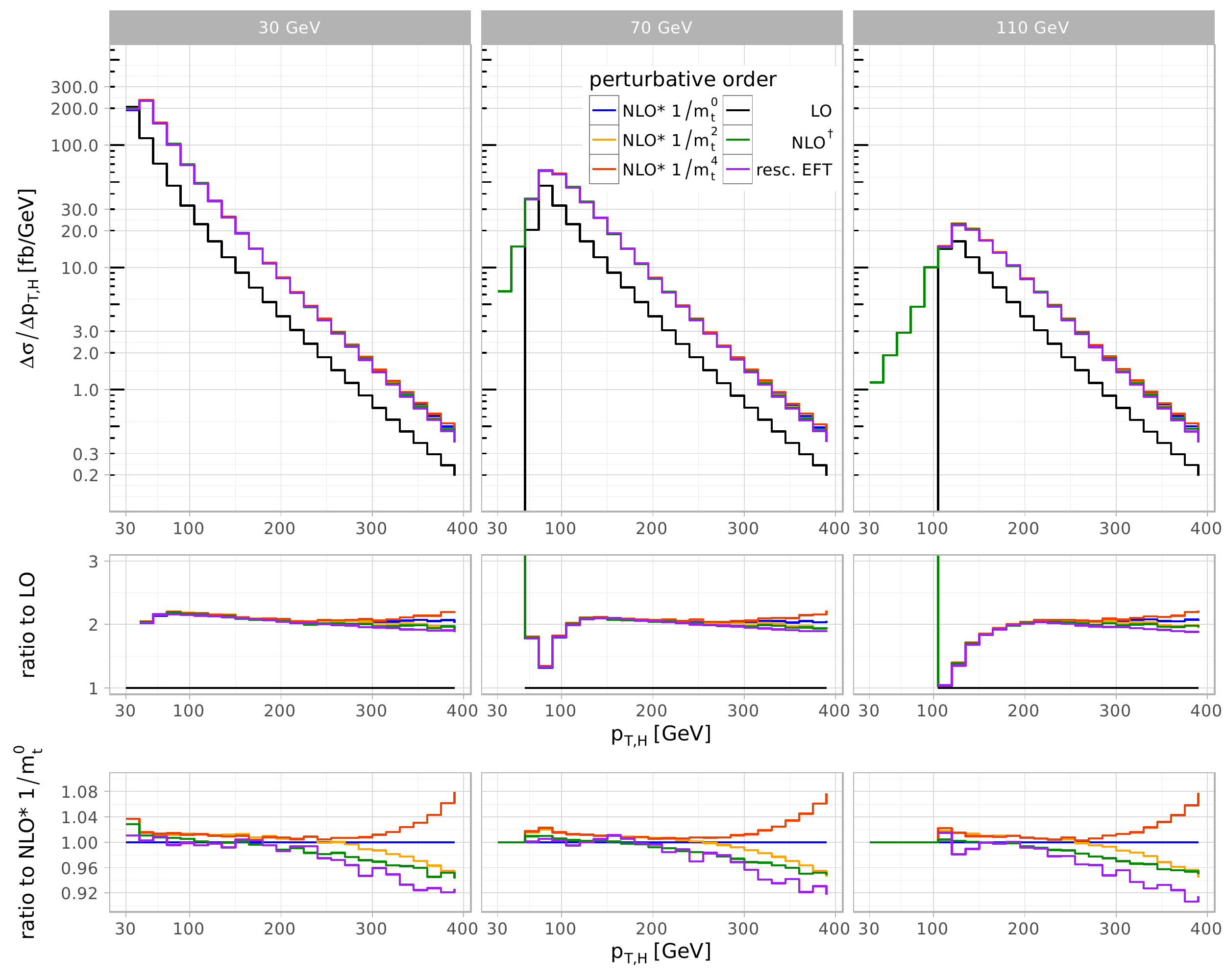}
	\caption{\label{fig:higgsptjetcut}\NLOs{} transverse momentum distributions of the Higgs boson with minimum jet transverse momenta of $30$, $70$ and \SI{110}{\GeV}. The upper panel shows the absolute distribution, while the lower two panels display the ratio to the \LO{} distribution and the \NLOs{} $1/m_t^0$ approximation, respectively.  }
\end{figure}

To consider a case where a \LO{} rescaling is not possible smoothly, we now study the Higgs $p_T$ distribution with an additional minimum jet transverse momentum requirement $p_{T,\text{jet}}^\text{min}$ in \cref{fig:higgsptjetcut}. Each column of the plot shows a different jet cut $p_{T,\text{jet}}^\text{min}$ of $30$,$70$ and $\SI{110}{GeV}$, respectively. For comparison the \NLOd{} approach is included again. Additionally, the \LO{} rescaled full \EFT{} result is included in purple. Note that for Higgs $p_T$ smaller than the minimum jet transverse momentum $p_{T,\text{jet}}^\text{min}$ this rescaling breaks down, or is rather ill defined.
Comparing both rescaling approaches with the \NLOs{} prediction, we observe that they underestimate the mass effects by a few percent at high $p_T$. For \SI{400}{GeV} the two rescaling approaches differ already by $4\%$, being comparable in size to the \NNLO{} scale uncertainty. Again we recommend using the $\mathcal{O}(m_t^{-2})$ terms for our \NLOs{} prediction with the same reasoning as above.

\begin{figure}
	\centering\includegraphics[width=0.8\textwidth]{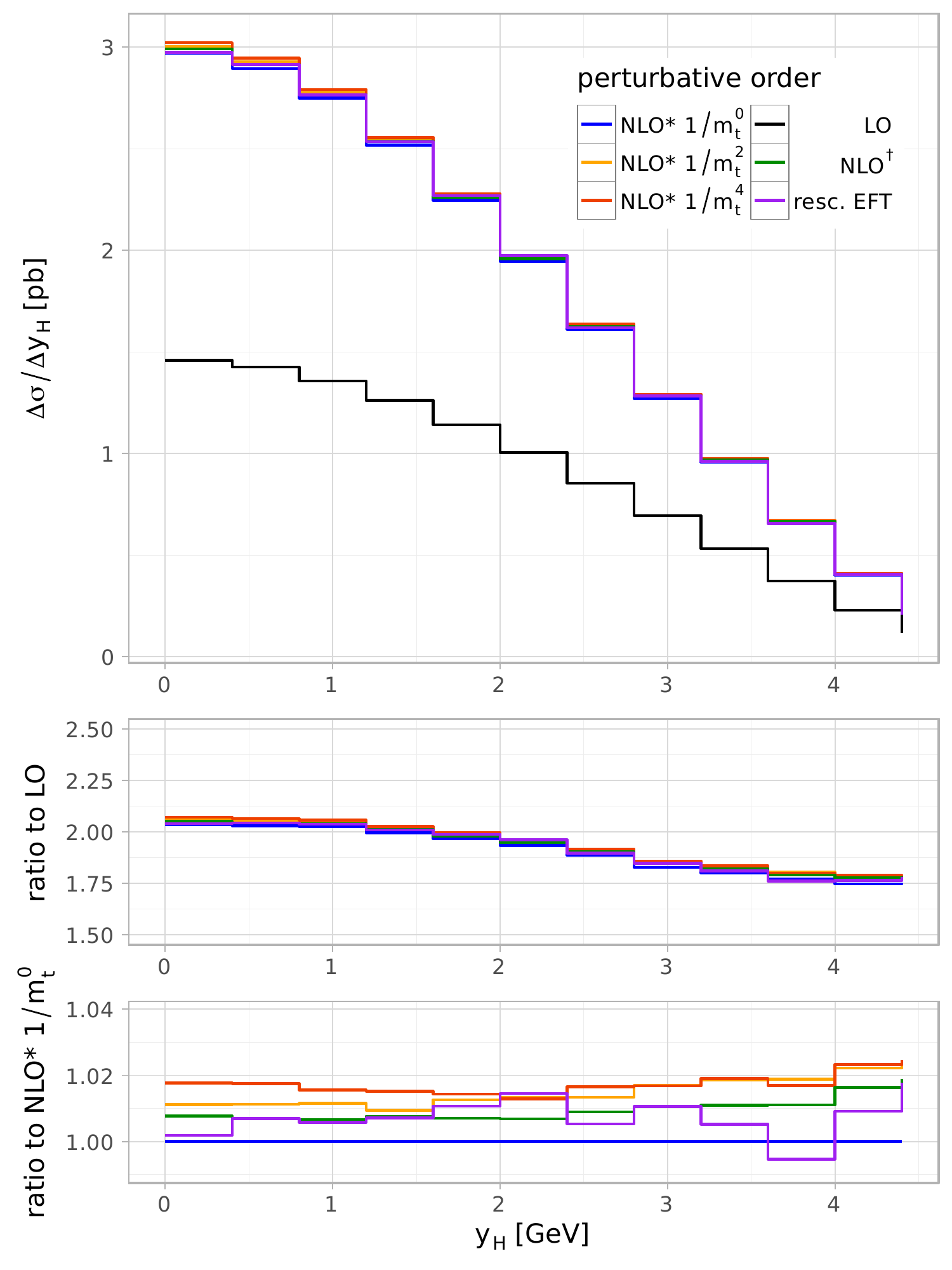}
	\caption{Higgs rapidity distribution for different orders and approximations. The upper panel shows the absolute distribution, while the lower two panels display the ratio to the \LO{} distribution and the {\abbrev NLO*} $1/m_t^0$ approximation, respectively.}
	\label{fig:etarap}
\end{figure}

Finally, to see how far our \NLOs{} result can improve predictions for $p_T$ inclusive observables we consider the Higgs rapidity distribution in \cref{fig:etarap}. Throughout the whole range of rapidities the higher orders of the asymptotic expansion coincide within 0.5\%. For large rapidities they coincide within the numerical uncertainty. Comparing this with the approach of \LO{} rescaled virtual corrections {\abbrev NLO$^\dagger$}, we observe a discrepancy of  0.5\% to 1\%.

\subsection{Higgs+jet NLO* phenomenology}

In the previous subsection we have shown different approximations for Higgs+jet observables which take into account top-mass effects and discussed their limitations. We argued taking the $\mathcal{O}(m_t^{-2})$ mass corrections into account for our \NLOs{} approximation as to get finite top-mass predictions up to scales of $\simeq\SI{250}{GeV}$ and reasonable predictions mimicking the rescaling behavior at higher energies. At energies above $\SI{300}{GeV}$ the remaining dependence on the asymptotic series becomes comparable in size to the residual \NNLO{} scale uncertainty of $\simeq8\%$ \cite{Chen:2016zka,Boughezal:2015aha,Boughezal:2013uia}.

\begin{figure}
	\centering\includegraphics[width=\textwidth]{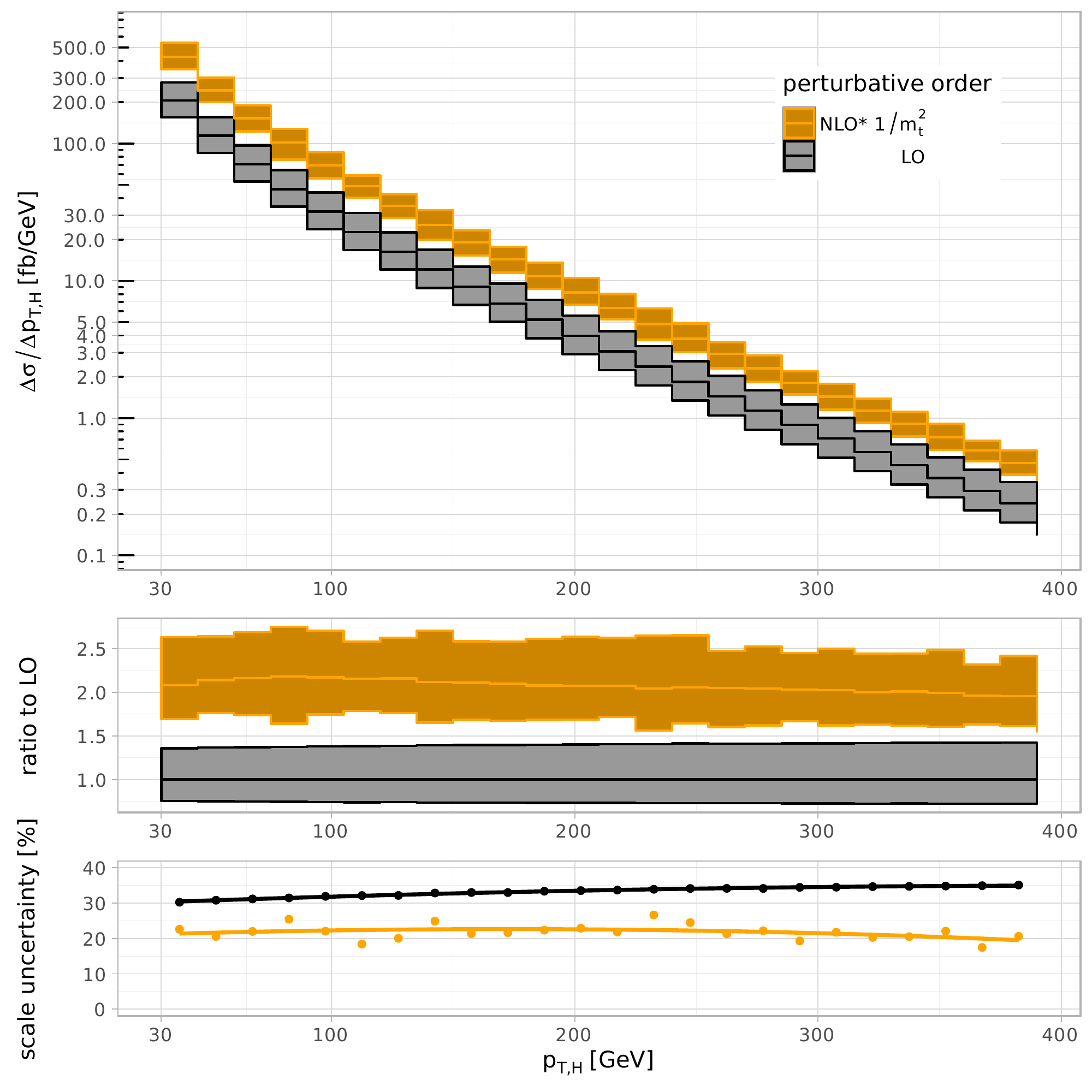}
	\caption{Higgs inclusive transverse momentum distribution at \LO{} and \NLOs{}. The ribbon is obtained by varying the factorization and renormalization scale $\mu=\sqrt{p_{T,H}^2 + m_H^2}$ by a factor of $2$ and $1/2$. The middle panel shows the ratio to the \LO{} cross section. The lower panel shows the mean of the upper and lower bound change with respect to the central value in percent, commonly known as the scale uncertainty.}
	\label{fig:pthscalevar}
\end{figure}

\begin{figure}
	\centering\includegraphics[width=0.8\textwidth]{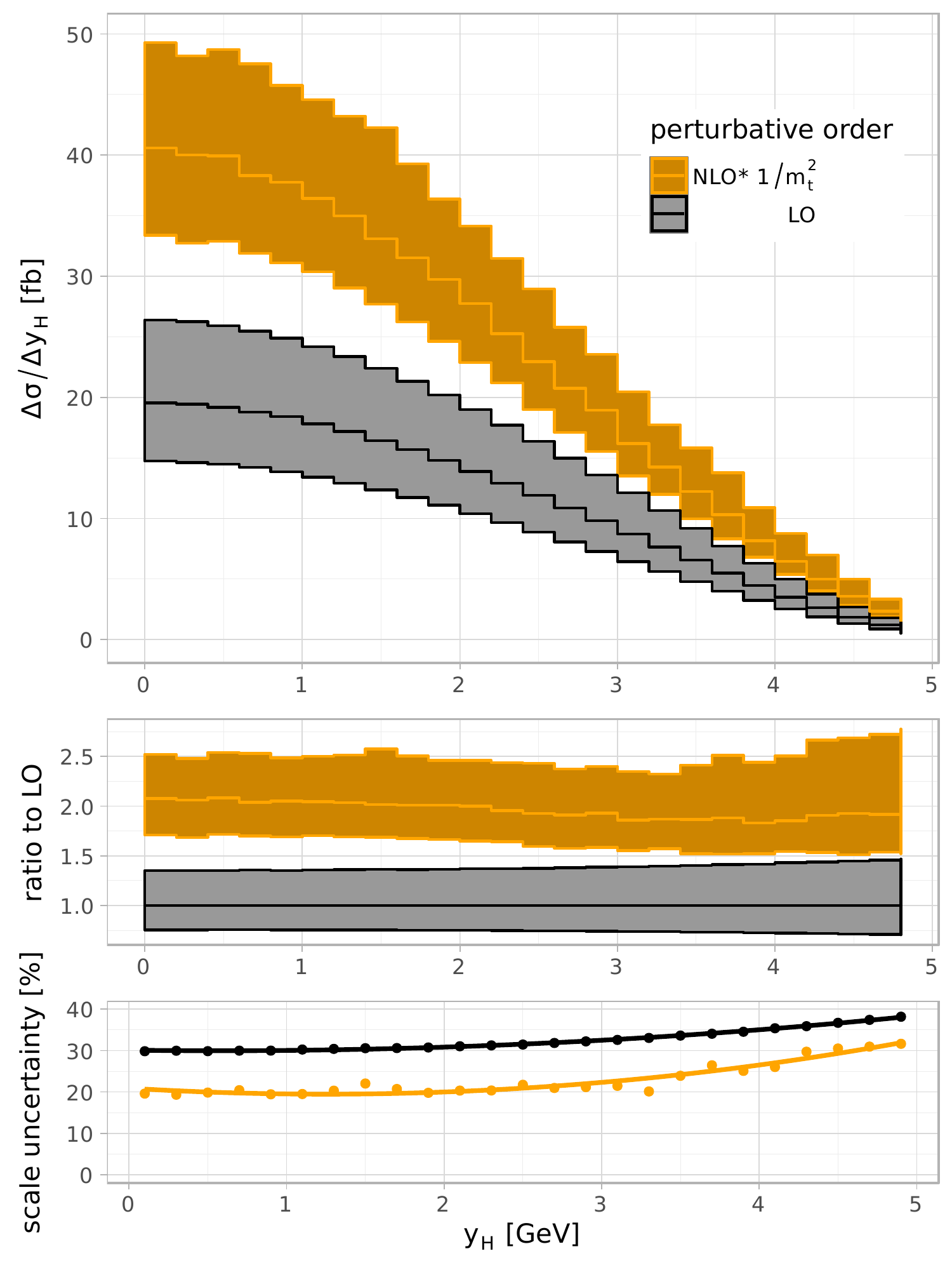}
	\caption{Higgs rapidity distribution at \LO{} and \NLOs{}. The ribbon is obtained by varying the factorization and renormalization scale $\mu=\sqrt{p_{T,H}^2 + m_H^2}$ by a factor of $2$ and $1/2$. The middle panel shows the ratio to the LO cross section. The lower panel shows the mean of the upper and lower bound change with respect to the central value in percent, commonly known as the scale uncertainty.}
	\label{fig:etarapmu}
\end{figure}

To estimate the dependence on the perturbative higher order corrections for the Higgs inclusive $p_T$ distribution we show the variation of the factorization and renormalization scale  $\mu_F=\mu_R=\mu=\sqrt{p_{T,H}^2 + m_H^2}$ by a factor of $2$ and $1/2$ in \cref{fig:pthscalevar}. The change of the central value by $\simeq20\%$ for the \NLOs{} cross section and $\simeq35\%$ for the \LO{} cross section confirms earlier results \cite{deFlorian:1999zd}. Similarly, we show the scale variation results for the Higgs rapdidity distribution in \cref{fig:etarapmu}. To obtain the best prediction for $H+\text{jet}$ observables we recommend taking the \NNLO{} \EFT{} results \cite{Chen:2016zka,Boughezal:2015aha,Boughezal:2013uia} and rescaling them by our \NLOs{}/\NLO{}-\EFT{} $K$-factor to minimize the scale dependence \emph{and} top-mass uncertainty.

\section{Conclusions} \label{sec:conc}

We presented a calculation of H+jet production at \NLO{} accuracy with full top-mass dependence in all components as far as possible with current technology: only the finite part of two-loop amplitude entering the virtual corrections is taken in an asymptotic expansion in $\Lambda/m_t$, where $\Lambda$ can be any kinematical scale of the process. The convergence of the asymptotic expansion allowed us to predict the full top-mass dependence for transverse momenta smaller than $\simeq\SI{225}{GeV}$ and quantify a remaining uncertainty for higher energies.

Previous approaches to include top-mass effects for H+jet cross sections were, for example, to rescale the \EFT{} result by the ratio of the $m_t$-exact to the \EFT{}  cross section at \LO{}. From a different point of view this equates to using the \EFT{} $K$-factor for the $m_t$-exact \LO{}. Further approximations have been made, taking the two parton real emission $m_t$-exactly into account and \LO{} rescaling the fully \EFT{} virtual corrections.

We validated the use of these \LO{} rescaling schemes to a few percent for Higgs $p_T$ distributions up to $\simeq\SI{250}{GeV}$ and provide $m_t$ exact predictions using the higher order corrections $1/m_t^2$, $1/m_t^4$ for scales up to $\simeq\SI{225}{GeV}$. For energies larger than $\simeq\SI{300}{GeV}$ the estimated error made by using the \LO{} rescaling scheme becomes comparable in size to the \NNLO{} scale uncertainty of $\simeq 8\%$ and a full calculation of the two-loop integrals is advised. We recommend using our \NLOs{} $\mathcal{O}(1/m_t^2)$ corrections to make predictions over the full kinematical range: We believe they are very close to the $m_t$-exact results for $p_T\lesssim\SI{225}{GeV}$ and mimic the \LO{} rescaled virtual corrections scheme for larger $p_T$. Alternatively one could consider a combination with the $m_t$-exact high $p_T$ limit in ref.~\cite{Caola:2016upw}. For sufficiently exclusive cuts, where no mapping to a born phase space is possible, an \NLO{} rescaling, using for example our \NLOs{} result, is needed.


To conclude, our \NLOs{} approximation for Higgs+jet production gives top-mass improved results up to high kinematical scales of $\simeq\SI{300}{GeV}$ relevant for \LHC{} Run II by including $m_t$-exact parts as far as currently possible and by exploiting asymptotic expansions for the remaining two-loop integrals. Our work facilitates the construction of the best fixed order Higgs+jet observables, by rescaling the \NNLO{}-\EFT{} results with our \NLOs{}/\NLO{}-\EFT{} top-mass corrections $K$-factor.

\paragraph*{Acknowledgments}
We are grateful to Stefano Carrazza for providing us with an advanced copy of {\abbrev QCDLoop} 2.0. We would like to thank Tom Zirke for help with the exp/q2e-setup and Robert Harlander and Frank Petriello for useful comments. Support was provided by the Center for Computational Research at the University at Buffalo. CW is supported by the National Science Foundation through award number PHY-1619877.

\bibliographystyle{JHEP}
\bibliography{HHpt}

\providecommand{\href}[2]{#2}\begingroup\raggedright\begin{thebibliography}{10}

\bibitem{Chatrchyan:2012xdj}
{\bf CMS} Collaboration, S.~Chatrchyan et~al., {\it {Observation of a new boson
  at a mass of 125 GeV with the CMS experiment at the LHC}},  {\em Phys. Lett.}
  {\bf B716} (2012) 30--61, [\href{http://arxiv.org/abs/1207.7235}{{\tt
  arXiv:1207.7235}}].

\bibitem{Aad:2012tfa}
{\bf ATLAS} Collaboration, G.~Aad et~al., {\it {Observation of a new particle
  in the search for the Standard Model Higgs boson with the ATLAS detector at
  the LHC}},  {\em Phys. Lett.} {\bf B716} (2012) 1--29,
  [\href{http://arxiv.org/abs/1207.7214}{{\tt arXiv:1207.7214}}].

\bibitem{Aad:2015lha}
{\bf ATLAS} Collaboration, G.~Aad et~al., {\it {Measurements of the Total and
  Differential Higgs Boson Production Cross Sections Combining the
  $H\ensuremath{\rightarrow}\ensuremath{\gamma}\ensuremath{\gamma}$ and
  $H\ensuremath{\rightarrow}Z{Z}^{*}\ensuremath{\rightarrow}4\ensuremath{\ell}$
  Decay Channels at $\sqrt{s}=8\text{ }\text{ }\mathrm{TeV}$ with the ATLAS
  Detector}},  {\em Phys. Rev. Lett.} {\bf 115} (2015), no.~9 091801,
  [\href{http://arxiv.org/abs/1504.0583}{{\tt arXiv:1504.0583}}].

\bibitem{ATLAS:2015bea}
{\bf ATLAS} Collaboration, T.~A. collaboration, {\it {Measurements of the Higgs
  boson production and decay rates and coupling strengths using pp collision
  data at $\sqrt{s}$ = 7 and 8 TeV in the ATLAS experiment}}, .

\bibitem{Aad:2015mxa}
{\bf ATLAS} Collaboration, G.~Aad et~al., {\it {Study of the spin and parity of
  the Higgs boson in diboson decays with the ATLAS detector}},  {\em Eur. Phys.
  J.} {\bf C75} (2015), no.~10 476, [\href{http://arxiv.org/abs/1506.0566}{{\tt
  arXiv:1506.0566}}]. [Erratum: Eur. Phys. J.C76,no.3,152(2016)].

\bibitem{Khachatryan:2014jba}
{\bf CMS} Collaboration, V.~Khachatryan et~al., {\it {Precise determination of
  the mass of the Higgs boson and tests of compatibility of its couplings with
  the standard model predictions using proton collisions at 7 and 8 $\,\text
  {TeV}$}},  {\em Eur. Phys. J.} {\bf C75} (2015), no.~5 212,
  [\href{http://arxiv.org/abs/1412.8662}{{\tt arXiv:1412.8662}}].

\bibitem{Aad:2015gba}
{\bf ATLAS} Collaboration, G.~Aad et~al., {\it {Measurements of the Higgs boson
  production and decay rates and coupling strengths using pp collision data at
  $\sqrt{s}=7$ and 8 TeV in the ATLAS experiment}},  {\em Eur. Phys. J.} {\bf
  C76} (2016), no.~1 6, [\href{http://arxiv.org/abs/1507.0454}{{\tt
  arXiv:1507.0454}}].

\bibitem{Giardino:2013bma}
P.~P. Giardino, K.~Kannike, I.~Masina, M.~Raidal, and A.~Strumia, {\it {The
  universal Higgs fit}},  {\em JHEP} {\bf 05} (2014) 046,
  [\href{http://arxiv.org/abs/1303.3570}{{\tt arXiv:1303.3570}}].

\bibitem{Anastasiou:2015ema}
C.~Anastasiou, C.~Duhr, F.~Dulat, F.~Herzog, and B.~Mistlberger, {\it {Higgs
  Boson Gluon-Fusion Production in QCD at Three Loops}},  {\em Phys. Rev.
  Lett.} {\bf 114} (2015) 212001, [\href{http://arxiv.org/abs/1503.0605}{{\tt
  arXiv:1503.0605}}].

\bibitem{Anastasiou:2016cez}
C.~Anastasiou, C.~Duhr, F.~Dulat, E.~Furlan, T.~Gehrmann, F.~Herzog,
  A.~Lazopoulos, and B.~Mistlberger, {\it {High precision determination of the
  gluon fusion Higgs boson cross-section at the LHC}},  {\em JHEP} {\bf 05}
  (2016) 058, [\href{http://arxiv.org/abs/1602.0069}{{\tt arXiv:1602.0069}}].

\bibitem{Aad:2016lvc}
{\bf ATLAS} Collaboration, G.~Aad et~al., {\it {Measurement of fiducial
  differential cross sections of gluon-fusion production of Higgs bosons
  decaying to $WW^{\ast}{\rightarrow\,}e\nu\mu\nu$ with the ATLAS detector at
  $\sqrt{s}=8$ TeV}},  \href{http://arxiv.org/abs/1604.0299}{{\tt
  arXiv:1604.0299}}.

\bibitem{Aad:2015tna}
{\bf ATLAS} Collaboration, G.~Aad et~al., {\it {Constraints on non-Standard
  Model Higgs boson interactions in an effective Lagrangian using differential
  cross sections measured in the $H \rightarrow \gamma\gamma$ decay channel at
  $\sqrt{s} = 8$TeV with the ATLAS detector}},  {\em Phys. Lett.} {\bf B753}
  (2016) 69--85, [\href{http://arxiv.org/abs/1508.0250}{{\tt
  arXiv:1508.0250}}].

\bibitem{Aad:2014tca}
{\bf ATLAS} Collaboration, G.~Aad et~al., {\it {Fiducial and differential cross
  sections of Higgs boson production measured in the four-lepton decay channel
  in $pp$ collisions at $\sqrt{s}$=8 TeV with the ATLAS detector}},  {\em Phys.
  Lett.} {\bf B738} (2014) 234--253,
  [\href{http://arxiv.org/abs/1408.3226}{{\tt arXiv:1408.3226}}].

\bibitem{Aad:2014lwa}
{\bf ATLAS} Collaboration, G.~Aad et~al., {\it {Measurements of fiducial and
  differential cross sections for Higgs boson production in the diphoton decay
  channel at $\sqrt{s}=8$ TeV with ATLAS}},  {\em JHEP} {\bf 09} (2014) 112,
  [\href{http://arxiv.org/abs/1407.4222}{{\tt arXiv:1407.4222}}].

\bibitem{Khachatryan:2015yvw}
{\bf CMS} Collaboration, V.~Khachatryan et~al., {\it {Measurement of
  differential and integrated fiducial cross sections for Higgs boson
  production in the four-lepton decay channel in pp collisions at $ \sqrt{s}=7
  $ and 8 TeV}},  {\em JHEP} {\bf 04} (2016) 005,
  [\href{http://arxiv.org/abs/1512.0837}{{\tt arXiv:1512.0837}}].

\bibitem{Khachatryan:2015rxa}
{\bf CMS} Collaboration, V.~Khachatryan et~al., {\it {Measurement of
  differential cross sections for Higgs boson production in the diphoton decay
  channel in pp collisions at $\sqrt{s}=8\,\text {TeV} $}},  {\em Eur. Phys.
  J.} {\bf C76} (2016), no.~1 13, [\href{http://arxiv.org/abs/1508.0781}{{\tt
  arXiv:1508.0781}}].

\bibitem{CMS:2015obs}
{\bf CMS} Collaboration, V.~Khachatryan et~al., {\it {Measurement of the
  transverse momentum spectrum of the Higgs boson produced in pp collisions at
  $\sqrt{s} = 8~\mathrm{TeV}$ using the H$\to$WW decays}}, .

\bibitem{Harlander:2013oja}
R.~V. Harlander and T.~Neumann, {\it {Probing the nature of the Higgs-gluon
  coupling}},  {\em Phys. Rev.} {\bf D88} (2013) 074015,
  [\href{http://arxiv.org/abs/1308.2225}{{\tt arXiv:1308.2225}}].

\bibitem{Dawson:2014ora}
S.~Dawson, I.~M. Lewis, and M.~Zeng, {\it {Effective field theory for Higgs
  boson plus jet production}},  {\em Phys. Rev.} {\bf D90} (2014), no.~9
  093007, [\href{http://arxiv.org/abs/1409.6299}{{\tt arXiv:1409.6299}}].

\bibitem{Dawson:2015gka}
S.~Dawson, I.~M. Lewis, and M.~Zeng, {\it {Usefulness of effective field theory
  for boosted Higgs production}},  {\em Phys. Rev.} {\bf D91} (2015) 074012,
  [\href{http://arxiv.org/abs/1501.0410}{{\tt arXiv:1501.0410}}].

\bibitem{Banfi:2013yoa}
A.~Banfi, A.~Martin, and V.~Sanz, {\it {Probing top-partners in Higgs+jets}},
  {\em JHEP} {\bf 08} (2014) 053, [\href{http://arxiv.org/abs/1308.4771}{{\tt
  arXiv:1308.4771}}].

\bibitem{Azatov:2013xha}
A.~Azatov and A.~Paul, {\it {Probing Higgs couplings with high $p_T$ Higgs
  production}},  {\em JHEP} {\bf 01} (2014) 014,
  [\href{http://arxiv.org/abs/1309.5273}{{\tt arXiv:1309.5273}}].

\bibitem{Grojean:2013nya}
C.~Grojean, E.~Salvioni, M.~Schlaffer, and A.~Weiler, {\it {Very boosted Higgs
  in gluon fusion}},  {\em JHEP} {\bf 05} (2014) 022,
  [\href{http://arxiv.org/abs/1312.3317}{{\tt arXiv:1312.3317}}].

\bibitem{Schlaffer:2014osa}
M.~Schlaffer, M.~Spannowsky, M.~Takeuchi, A.~Weiler, and C.~Wymant, {\it
  {Boosted Higgs Shapes}},  {\em Eur. Phys. J.} {\bf C74} (2014), no.~10 3120,
  [\href{http://arxiv.org/abs/1405.4295}{{\tt arXiv:1405.4295}}].

\bibitem{Buschmann:2014twa}
M.~Buschmann, C.~Englert, D.~Goncalves, T.~Plehn, and M.~Spannowsky, {\it
  {Resolving the Higgs-Gluon Coupling with Jets}},  {\em Phys. Rev.} {\bf D90}
  (2014), no.~1 013010, [\href{http://arxiv.org/abs/1405.7651}{{\tt
  arXiv:1405.7651}}].

\bibitem{Buschmann:2014sia}
M.~Buschmann, D.~Goncalves, S.~Kuttimalai, M.~Sch\"onherr, F.~Krauss, and
  T.~Plehn, {\it {Mass Effects in the Higgs-Gluon Coupling: Boosted vs
  Off-Shell Production}},  {\em JHEP} {\bf 02} (2015) 038,
  [\href{http://arxiv.org/abs/1410.5806}{{\tt arXiv:1410.5806}}].

\bibitem{Langenegger:2015lra}
U.~Langenegger, M.~Spira, and I.~Strebel, {\it {Testing the Higgs Boson
  Coupling to Gluons}},  \href{http://arxiv.org/abs/1507.0137}{{\tt
  arXiv:1507.0137}}.

\bibitem{Ghosh:2014wxa}
D.~Ghosh and M.~Wiebusch, {\it {Dimension-six triple gluon operator in
  Higgs$+$jet observables}},  {\em Phys. Rev.} {\bf D91} (2015), no.~3 031701,
  [\href{http://arxiv.org/abs/1411.2029}{{\tt arXiv:1411.2029}}].

\bibitem{Boughezal:2013uia}
R.~Boughezal, F.~Caola, K.~Melnikov, F.~Petriello, and M.~Schulze, {\it {Higgs
  boson production in association with a jet at next-to-next-to-leading order
  in perturbative QCD}},  {\em JHEP} {\bf 06} (2013) 072,
  [\href{http://arxiv.org/abs/1302.6216}{{\tt arXiv:1302.6216}}].

\bibitem{Chen:2014gva}
X.~Chen, T.~Gehrmann, E.~W.~N. Glover, and M.~Jaquier, {\it {Precise QCD
  predictions for the production of Higgs + jet final states}},  {\em Phys.
  Lett.} {\bf B740} (2015) 147--150,
  [\href{http://arxiv.org/abs/1408.5325}{{\tt arXiv:1408.5325}}].

\bibitem{Boughezal:2015dra}
R.~Boughezal, F.~Caola, K.~Melnikov, F.~Petriello, and M.~Schulze, {\it {Higgs
  boson production in association with a jet at next-to-next-to-leading
  order}},  {\em Phys. Rev. Lett.} {\bf 115} (2015), no.~8 082003,
  [\href{http://arxiv.org/abs/1504.0792}{{\tt arXiv:1504.0792}}].

\bibitem{Boughezal:2015aha}
R.~Boughezal, C.~Focke, W.~Giele, X.~Liu, and F.~Petriello, {\it {Higgs boson
  production in association with a jet at NNLO using jettiness subtraction}},
  {\em Phys. Lett.} {\bf B748} (2015) 5--8,
  [\href{http://arxiv.org/abs/1505.0389}{{\tt arXiv:1505.0389}}].

\bibitem{Campbell:2006xx}
J.~M. Campbell, R.~K. Ellis, and G.~Zanderighi, {\it {Next-to-Leading order
  Higgs + 2 jet production via gluon fusion}},  {\em JHEP} {\bf 10} (2006) 028,
  [\href{http://arxiv.org/abs/hep-ph/0608194}{{\tt hep-ph/0608194}}].

\bibitem{Campbell:2010cz}
J.~M. Campbell, R.~K. Ellis, and C.~Williams, {\it {Hadronic production of a
  Higgs boson and two jets at next-to-leading order}},  {\em Phys. Rev.} {\bf
  D81} (2010) 074023, [\href{http://arxiv.org/abs/1001.4495}{{\tt
  arXiv:1001.4495}}].

\bibitem{Cullen:2013saa}
G.~Cullen, H.~van Deurzen, N.~Greiner, G.~Luisoni, P.~Mastrolia, E.~Mirabella,
  G.~Ossola, T.~Peraro, and F.~Tramontano, {\it {Next-to-Leading-Order QCD
  Corrections to Higgs Boson Production Plus Three Jets in Gluon Fusion}},
  {\em Phys. Rev. Lett.} {\bf 111} (2013), no.~13 131801,
  [\href{http://arxiv.org/abs/1307.4737}{{\tt arXiv:1307.4737}}].

\bibitem{Graudenz:1992pv}
D.~Graudenz, M.~Spira, and P.~M. Zerwas, {\it {QCD corrections to Higgs boson
  production at proton proton colliders}},  {\em Phys. Rev. Lett.} {\bf 70}
  (1993) 1372--1375.

\bibitem{Harlander:2005rq}
R.~Harlander and P.~Kant, {\it {Higgs production and decay: Analytic results at
  next-to-leading order QCD}},  {\em JHEP} {\bf 12} (2005) 015,
  [\href{http://arxiv.org/abs/hep-ph/0509189}{{\tt hep-ph/0509189}}].

\bibitem{Anastasiou:2006hc}
C.~Anastasiou, S.~Beerli, S.~Bucherer, A.~Daleo, and Z.~Kunszt, {\it {Two-loop
  amplitudes and master integrals for the production of a Higgs boson via a
  massive quark and a scalar-quark loop}},  {\em JHEP} {\bf 01} (2007) 082,
  [\href{http://arxiv.org/abs/hep-ph/0611236}{{\tt hep-ph/0611236}}].

\bibitem{Harlander:2009mq}
R.~V. Harlander and K.~J. Ozeren, {\it {Finite top mass effects for hadronic
  Higgs production at next-to-next-to-leading order}},  {\em JHEP} {\bf 11}
  (2009) 088, [\href{http://arxiv.org/abs/0909.3420}{{\tt arXiv:0909.3420}}].

\bibitem{Pak:2009dg}
A.~Pak, M.~Rogal, and M.~Steinhauser, {\it {Finite top quark mass effects in
  NNLO Higgs boson production at LHC}},  {\em JHEP} {\bf 02} (2010) 025,
  [\href{http://arxiv.org/abs/0911.4662}{{\tt arXiv:0911.4662}}].

\bibitem{Harlander:2009bw}
R.~V. Harlander and K.~J. Ozeren, {\it {Top mass effects in Higgs production at
  next-to-next-to-leading order QCD: Virtual corrections}},  {\em Phys. Lett.}
  {\bf B679} (2009) 467--472, [\href{http://arxiv.org/abs/0907.2997}{{\tt
  arXiv:0907.2997}}].

\bibitem{Pak:2009bx}
A.~Pak, M.~Rogal, and M.~Steinhauser, {\it {Virtual three-loop corrections to
  Higgs boson production in gluon fusion for finite top quark mass}},  {\em
  Phys. Lett.} {\bf B679} (2009) 473--477,
  [\href{http://arxiv.org/abs/0907.2998}{{\tt arXiv:0907.2998}}].

\bibitem{Harlander:2016hcx}
R.~V. Harlander, S.~Liebler, and H.~Mantler, {\it {SusHi Bento: Beyond NNLO and
  the heavy-top limit}},  \href{http://arxiv.org/abs/1605.0319}{{\tt
  arXiv:1605.0319}}.

\bibitem{Marzani:2008ih}
S.~Marzani, R.~D. Ball, V.~Del~Duca, S.~Forte, and A.~Vicini, {\it
  {Finite-top-mass effects in NNLO Higgs production}},  {\em Nucl. Phys. Proc.
  Suppl.} {\bf 186} (2009) 98--101, [\href{http://arxiv.org/abs/0809.4934}{{\tt
  arXiv:0809.4934}}].

\bibitem{Marzani:2008az}
S.~Marzani, R.~D. Ball, V.~Del~Duca, S.~Forte, and A.~Vicini, {\it {Higgs
  production via gluon-gluon fusion with finite top mass beyond next-to-leading
  order}},  {\em Nucl. Phys.} {\bf B800} (2008) 127--145,
  [\href{http://arxiv.org/abs/0801.2544}{{\tt arXiv:0801.2544}}].

\bibitem{Harlander:2009my}
R.~V. Harlander, H.~Mantler, S.~Marzani, and K.~J. Ozeren, {\it {Higgs
  production in gluon fusion at next-to-next-to-leading order QCD for finite
  top mass}},  {\em Eur. Phys. J.} {\bf C66} (2010) 359--372,
  [\href{http://arxiv.org/abs/0912.2104}{{\tt arXiv:0912.2104}}].

\bibitem{DelDuca:2003ba}
V.~Del~Duca, W.~Kilgore, C.~Oleari, C.~R. Schmidt, and D.~Zeppenfeld, {\it
  {Kinematical limits on Higgs boson production via gluon fusion in association
  with jets}},  {\em Phys. Rev.} {\bf D67} (2003) 073003,
  [\href{http://arxiv.org/abs/hep-ph/0301013}{{\tt hep-ph/0301013}}].

\bibitem{Baur:1989cm}
U.~Baur and E.~W.~N. Glover, {\it {Higgs Boson Production at Large Transverse
  Momentum in Hadronic Collisions}},  {\em Nucl. Phys.} {\bf B339} (1990)
  38--66.

\bibitem{Ellis:1987xu}
R.~K. Ellis, I.~Hinchliffe, M.~Soldate, and J.~J. van~der Bij, {\it {Higgs
  Decay to tau+ tau-: A Possible Signature of Intermediate Mass Higgs Bosons at
  the SSC}},  {\em Nucl. Phys.} {\bf B297} (1988) 221--243.

\bibitem{DelDuca:2001eu}
V.~Del~Duca, W.~Kilgore, C.~Oleari, C.~Schmidt, and D.~Zeppenfeld, {\it {Higgs
  + 2 jets via gluon fusion}},  {\em Phys. Rev. Lett.} {\bf 87} (2001) 122001,
  [\href{http://arxiv.org/abs/hep-ph/0105129}{{\tt hep-ph/0105129}}].

\bibitem{DelDuca:2001fn}
V.~Del~Duca, W.~Kilgore, C.~Oleari, C.~Schmidt, and D.~Zeppenfeld, {\it {Gluon
  fusion contributions to H + 2 jet production}},  {\em Nucl. Phys.} {\bf B616}
  (2001) 367--399, [\href{http://arxiv.org/abs/hep-ph/0108030}{{\tt
  hep-ph/0108030}}].

\bibitem{Campanario:2013mga}
F.~Campanario and M.~Kubocz, {\it {Higgs boson production in association with
  three jets via gluon fusion at the LHC: Gluonic contributions}},  {\em Phys.
  Rev.} {\bf D88} (2013), no.~5 054021,
  [\href{http://arxiv.org/abs/1306.1830}{{\tt arXiv:1306.1830}}].

\bibitem{Greiner:2016awe}
N.~Greiner, S.~Hoeche, G.~Luisoni, M.~Sch\"onherr, and J.-C. Winter, {\it {Full
  mass dependence in Higgs boson production in association with jets at the LHC
  and FCC}},  \href{http://arxiv.org/abs/1608.0119}{{\tt arXiv:1608.0119}}.

\bibitem{Frederix:2016cnl}
R.~Frederix, S.~Frixione, E.~Vryonidou, and M.~Wiesemann, {\it {Heavy-quark
  mass effects in Higgs plus jets production}},
  \href{http://arxiv.org/abs/1604.0301}{{\tt arXiv:1604.0301}}.

\bibitem{Borowka:2016ypz}
S.~Borowka, N.~Greiner, G.~Heinrich, S.~P. Jones, M.~Kerner, J.~Schlenk, and
  T.~Zirke, {\it {Full top quark mass dependence in Higgs boson pair production
  at NLO}},  \href{http://arxiv.org/abs/1608.0479}{{\tt arXiv:1608.0479}}.

\bibitem{Borowka:2016ehy}
S.~Borowka, N.~Greiner, G.~Heinrich, S.~Jones, M.~Kerner, J.~Schlenk,
  U.~Schubert, and T.~Zirke, {\it {Higgs Boson Pair Production in Gluon Fusion
  at Next-to-Leading Order with Full Top-Quark Mass Dependence}},  {\em Phys.
  Rev. Lett.} {\bf 117} (2016), no.~1 012001,
  [\href{http://arxiv.org/abs/1604.0644}{{\tt arXiv:1604.0644}}]. [Erratum:
  Phys. Rev. Lett.117,no.7,079901(2016)].

\bibitem{Forte:2015gve}
S.~Forte and C.~Muselli, {\it {High energy resummation of transverse momentum
  distributions: Higgs in gluon fusion}},  {\em JHEP} {\bf 03} (2016) 122,
  [\href{http://arxiv.org/abs/1511.0556}{{\tt arXiv:1511.0556}}].

\bibitem{Caola:2016upw}
F.~Caola, S.~Forte, S.~Marzani, C.~Muselli, and G.~Vita, {\it {The Higgs
  transverse momentum spectrum with finite quark masses beyond leading order}},
   \href{http://arxiv.org/abs/1606.0410}{{\tt arXiv:1606.0410}}.

\bibitem{Neumann:2014nha}
T.~Neumann and M.~Wiesemann, {\it {Finite top-mass effects in gluon-induced
  Higgs production with a jet-veto at NNLO}},  {\em JHEP} {\bf 1411} (2014)
  150, [\href{http://arxiv.org/abs/1408.6836}{{\tt arXiv:1408.6836}}].

\bibitem{Harlander:2012hf}
R.~V. Harlander, T.~Neumann, K.~J. Ozeren, and M.~Wiesemann, {\it {Top-mass
  effects in differential Higgs production through gluon fusion at order
  $\alpha_s^4$}},  {\em JHEP} {\bf 1208} (2012) 139,
  [\href{http://arxiv.org/abs/1206.0157}{{\tt arXiv:1206.0157}}].

\bibitem{Campbell:1999ah}
J.~M. Campbell and R.~K. Ellis, {\it {An Update on vector boson pair production
  at hadron colliders}},  {\em Phys. Rev.} {\bf D60} (1999) 113006,
  [\href{http://arxiv.org/abs/hep-ph/9905386}{{\tt hep-ph/9905386}}].

\bibitem{Campbell:2011bn}
J.~M. Campbell, R.~K. Ellis, and C.~Williams, {\it {Vector boson pair
  production at the LHC}},  {\em JHEP} {\bf 07} (2011) 018,
  [\href{http://arxiv.org/abs/1105.0020}{{\tt arXiv:1105.0020}}].

\bibitem{Campbell:2015qma}
J.~M. Campbell, R.~K. Ellis, and W.~T. Giele, {\it {A Multi-Threaded Version of
  MCFM}},  {\em Eur. Phys. J.} {\bf C75} (2015), no.~6 246,
  [\href{http://arxiv.org/abs/1503.0618}{{\tt arXiv:1503.0618}}].

\bibitem{Mantler:2009}
H.~Mantler, {\it {Bottom-Quark-Effekte in der Higgs-Produktion}},  Master's
  thesis, University of Wuppertal, 2009.

\bibitem{Keung:2009bs}
W.-Y. Keung and F.~J. Petriello, {\it {Electroweak and finite quark-mass
  effects on the Higgs boson transverse momentum distribution}},  {\em Phys.
  Rev.} {\bf D80} (2009) 013007, [\href{http://arxiv.org/abs/0905.2775}{{\tt
  arXiv:0905.2775}}].

\bibitem{Harlander:2012pb}
R.~V. Harlander, S.~Liebler, and H.~Mantler, {\it {SusHi: A program for the
  calculation of Higgs production in gluon fusion and bottom-quark annihilation
  in the Standard Model and the MSSM}},  {\em Comput. Phys. Commun.} {\bf 184}
  (2013) 1605--1617, [\href{http://arxiv.org/abs/1212.3249}{{\tt
  arXiv:1212.3249}}].

\bibitem{Bonciani:2015eua}
R.~Bonciani, V.~Del~Duca, H.~Frellesvig, J.~M. Henn, F.~Moriello, and V.~A.
  Smirnov, {\it {Next-to-leading order QCD corrections to the decay width H
  $\to$ Z$\gamma$}},  {\em JHEP} {\bf 08} (2015) 108,
  [\href{http://arxiv.org/abs/1505.0056}{{\tt arXiv:1505.0056}}].

\bibitem{Catani:1996vz}
S.~Catani and M.~H. Seymour, {\it {A General algorithm for calculating jet
  cross-sections in NLO QCD}},  {\em Nucl. Phys.} {\bf B485} (1997) 291--419,
  [\href{http://arxiv.org/abs/hep-ph/9605323}{{\tt hep-ph/9605323}}]. [Erratum:
  Nucl. Phys.B510,503(1998)].

\bibitem{Chen:2016zka}
X.~Chen, J.~Cruz-Martinez, T.~Gehrmann, E.~W.~N. Glover, and M.~Jaquier, {\it
  {NNLO QCD corrections to Higgs boson production at large transverse
  momentum}},  \href{http://arxiv.org/abs/1607.0881}{{\tt arXiv:1607.0881}}.

\bibitem{Harlander:1999cs}
R.~Harlander, {\it {Asymptotic expansions: Methods and applications}},  {\em
  Acta Phys. Polon.} {\bf B30} (1999) 3443--3462,
  [\href{http://arxiv.org/abs/hep-ph/9910496}{{\tt hep-ph/9910496}}].

\bibitem{Smirnov:2002pj}
V.~A. Smirnov, {\it {Applied asymptotic expansions in momenta and masses}},
  {\em Springer Tracts Mod. Phys.} {\bf 177} (2002) 1--262.

\bibitem{Badger:2008cm}
S.~D. Badger, {\it {Direct Extraction Of One Loop Rational Terms}},  {\em JHEP}
  {\bf 01} (2009) 049, [\href{http://arxiv.org/abs/0806.4600}{{\tt
  arXiv:0806.4600}}].

\bibitem{Mastrolia:2009dr}
P.~Mastrolia, {\it {Double-Cut of Scattering Amplitudes and Stokes' Theorem}},
  {\em Phys. Lett.} {\bf B678} (2009) 246--249,
  [\href{http://arxiv.org/abs/0905.2909}{{\tt arXiv:0905.2909}}].

\bibitem{Maitre:2007jq}
D.~Ma\^{\i}tre and P.~Mastrolia, {\it {S@M, a Mathematica Implementation of the
  Spinor-Helicity Formalism}},  {\em Comput. Phys. Commun.} {\bf 179} (2008)
  501--574, [\href{http://arxiv.org/abs/0710.5559}{{\tt arXiv:0710.5559}}].

\bibitem{Ellis:2008ir}
R.~K. Ellis, W.~T. Giele, Z.~Kunszt, and K.~Melnikov, {\it {Masses, fermions
  and generalized $D$-dimensional unitarity}},  {\em Nucl. Phys.} {\bf B822}
  (2009) 270--282, [\href{http://arxiv.org/abs/0806.3467}{{\tt
  arXiv:0806.3467}}].

\bibitem{Smirnov:1994tg}
V.~A. Smirnov, {\it {Asymptotic expansions in momenta and masses and
  calculation of Feynman diagrams}},  {\em Mod. Phys. Lett.} {\bf A10} (1995)
  1485--1500, [\href{http://arxiv.org/abs/hep-th/9412063}{{\tt
  hep-th/9412063}}].

\bibitem{Harlander:1997zb}
R.~Harlander, T.~Seidensticker, and M.~Steinhauser, {\it {Corrections of
  $\mathcal{O}(\alpha\alpha_s)$ to the decay of the $Z$ boson into bottom
  quarks}},  {\em Phys. Lett.} {\bf B426} (1998) 125--132,
  [\href{http://arxiv.org/abs/hep-ph/9712228}{{\tt hep-ph/9712228}}].

\bibitem{Seidensticker:1999bb}
T.~Seidensticker, {\it {Automatic application of successive asymptotic
  expansions of Feynman diagrams}},
  \href{http://arxiv.org/abs/hep-ph/9905298}{{\tt hep-ph/9905298}}.

\bibitem{Steinhauser:2000ry}
M.~Steinhauser, {\it {MATAD: A Program package for the computation of MAssive
  TADpoles}},  {\em Comput. Phys. Commun.} {\bf 134} (2001) 335--364,
  [\href{http://arxiv.org/abs/hep-ph/0009029}{{\tt hep-ph/0009029}}].

\bibitem{vonManteuffel:2012np}
A.~von Manteuffel and C.~Studerus, {\it {Reduze 2 - Distributed Feynman
  Integral Reduction}},  \href{http://arxiv.org/abs/1201.4330}{{\tt
  arXiv:1201.4330}}.

\bibitem{Ellis:2007qk}
R.~K. Ellis and G.~Zanderighi, {\it {Scalar one-loop integrals for QCD}},  {\em
  JHEP} {\bf 02} (2008) 002, [\href{http://arxiv.org/abs/0712.1851}{{\tt
  arXiv:0712.1851}}].

\bibitem{Carrazza:2016gav}
S.~Carrazza, R.~K. Ellis, and G.~Zanderighi, {\it {QCDLoop: a comprehensive
  framework for one-loop scalar integrals}},
  \href{http://arxiv.org/abs/1605.0318}{{\tt arXiv:1605.0318}}.

\bibitem{Dulat:2015mca}
S.~Dulat, T.-J. Hou, J.~Gao, M.~Guzzi, J.~Huston, P.~Nadolsky, J.~Pumplin,
  C.~Schmidt, D.~Stump, and C.~P. Yuan, {\it {New parton distribution functions
  from a global analysis of quantum chromodynamics}},  {\em Phys. Rev.} {\bf
  D93} (2016), no.~3 033006, [\href{http://arxiv.org/abs/1506.0744}{{\tt
  arXiv:1506.0744}}].

\bibitem{deFlorian:1999zd}
D.~de~Florian, M.~Grazzini, and Z.~Kunszt, {\it {Higgs production with large
  transverse momentum in hadronic collisions at next-to-leading order}},  {\em
  Phys. Rev. Lett.} {\bf 82} (1999) 5209--5212,
  [\href{http://arxiv.org/abs/hep-ph/9902483}{{\tt hep-ph/9902483}}].

\end{thebibliography}\endgroup


\providecommand{\href}[2]{#2}\begingroup\raggedright\endgroup

\end{document}